\documentclass[%
 reprint,
superscriptaddress,
caption
 amsmath,amssymb,
 aps,
prl,
]{revtex4-1}
\usepackage{xr}
\usepackage{graphicx}
\usepackage{dcolumn}
\usepackage{bm}
\usepackage{mathtools}
\usepackage{color,soul}
\usepackage{siunitx}
\usepackage{todonotes}
\usepackage{hyperref}
\usepackage{pdfpages}

\makeatletter
\AtBeginDocument{\let\LS@rot\@undefined}
\makeatother

\makeatletter
\newcommand*{\addFileDependency}[1]{
  \typeout{(#1)}
  \@addtofilelist{#1}
  \IfFileExists{#1}{}{\typeout{No file #1.}}
}
\makeatother

\newcommand*{\myexternaldocument}[1]{%
    \externaldocument{#1}%
    \addFileDependency{#1.tex}%
    \addFileDependency{#1.aux}%
}
\myexternaldocument{supplement}

\newcommand{\Pe}{\mathrm{Pe}}

\begin{document}

\title{Robust increase in supply by vessel dilation in globally coupled microvasculature}
\author{Felix J. Meigel}
\affiliation{Max Planck Institute for Dynamics and Self-Organization, 37077 G\"ottingen, Germany}
\author{Peter Cha}
\affiliation{John A.~Paulson School of Engineering and Applied Sciences and Kavli Institute for Bionano Science and Technology, Harvard University, Cambridge, MA 02138, USA}
\author{Michael P. Brenner}
\affiliation{John A.~Paulson School of Engineering and Applied Sciences and Kavli Institute for Bionano Science and Technology, Harvard University, Cambridge, MA 02138, USA}
\author{Karen Alim}
\affiliation{Max Planck Institute for Dynamics and Self-Organization, 37077 G\"ottingen, Germany}
\affiliation{Physik-Department, Technische Universit\"at M\"unchen, 85748 Garching, Germany}
\email{karen.alim@ds.mpg.de}
\begin{abstract}
Neuronal activity induces changes in blood flow by locally dilating vessels in the brain microvasculature. How can the local dilation of a single vessel increase flow-based metabolite supply, given that flows are globally coupled within microvasculature? Solving the supply dynamics for rat brain microvasculature, we find one parameter regime to dominate physiologically. This regime allows for robust increase in supply independent of the position in the network, which we explain analytically. We show that local coupling of vessels promotes spatially correlated increased supply by dilation.
\end{abstract}
\maketitle
Vascular networks pervade all organs of animals and are the paradigm of adaptive transport networks. Their self-organized architecture continuously inspires the search for their underlying physical principles \cite{Murray:1926tj,West:1997ww,Hu:2013io,Ronellenfitsch:2016hh,West:1997ww} and at the same time serves as a template for designing efficient networks in engineering \cite{Zheng:2017jg}. The blood flowing through vessels transports nutrients, hormones, and metabolites to adjacent tissues. Metabolite exchange primarily occurs within the fine vessel meshwork formed by microvasculature. In the brain, local metabolite demand can abruptly rise due to an increase in neural activity \cite{boynton_linear_1996},
 altering blood flow \cite{raichle_brain_2006,blinder_cortical_2013} in the same brain region, observable in fMRI \cite{logothetis_what_2008}. During the process of increased neuronal activity, neurons signal their increased demand to adjacent astrocyte cells, which in turn trigger small ring muscles surrounding blood vessels to relax \cite{macvicar_astrocyte_2015}. Thus, neural activity drives local dilation of a vessel \cite{cai_stimulation-induced_2018,hill_regional_2015}, and hence regulates metabolite supply \cite{vanzetta_compartment-resolved_2005,raichle_brain_2006}. However, from a fluid dynamics perspective there is a mystery:  blood vessels form a highly interconnected network in the microvasculature \cite{blinder_cortical_2013}, resulting in a global coupling of blood flow. A single dilating vessel can potentially change the metabolite supply in a broad region of the network -  and thus the local increase due to dilation is a function of specific network topology. Quantitatively, how much control over changes in blood-based supply resides in a single dilating vessel? 
  
Models considering metabolite spread in tissue date back more than a hundred years to A.~Krogh \cite{krogh_number_1919}. Krogh's model estimates the supply pattern in a tissue enclosed by vessels assuming that supply is constant on all vessel walls. Yet, on a larger tissue scale, supply spatially varies along the vasculature since resources supplied upstream are not available downstream. Alternative models consider vessel-based transport \cite{schneider_tissue_2012}, yet only diffusive transport is taken into account. The combined importance of advection and diffusion for transporting solutes in a \emph{single} tube was discovered by G.I. Taylor \cite{taylor_dispersion_1953,aris_dispersion_1956}, with subsequent work outlining modifications due to solute absorption at the tube boundary  \cite{e._m._lungu_effect_1982,shapiro_taylor_1986,mercer_complete_1994}. Yet, there has been much less work  capturing the coupling of advection and diffusion in tubular network structures \cite{marbach_pruning_2016,fang_oxygen_2008}, including solute absorption \cite{meigel_flow_2018}. The impact of a dilating vessel is hard to estimate since not only the absorption dynamics on the level of single vessels is changed, but also solute flux throughout the network is rerouted since fluid flow and thus solute flux are globally coupled. However, to connect fMRI, which relies on a fluid dynamic signal \cite{he_ultra-slow_2018,tian_cortical_2010,logothetis_what_2008}, and the change in blood flow with neuronal activity \cite{Paulson,cai_stimulation-induced_2018,raichle_brain_2006,oherron_neural_2016,peppiatt_bidirectional_2006}, we need to understand how vessel dilations affect the supply with metabolites.

In this letter, we present a theoretical model to determine the change in supply resulting from the dilation of a single vessel. On the level of an individual vessel, we analytically identify three regimes, each yielding a different functional dependence of the overall supply by absorption along the vessel wall on vessel geometry, blood flow, and blood flow based solute flux. Numerically analyzing supply dynamics in a microvasculature excerpt of a rat brain supplied from the Kleinfeld laboratory \cite{blinder_cortical_2013}, we find that a single regime dominates. This regime has the important property that dilating a single vessel robustly increases the supply along the dilated vessel independent of the exact location of the vessel in the network. We explain analytically how a single vessel can buffer the global coupling of solute fluxes within the network and yield a robust local increase independent of network topology. We further discuss how a single dilating vessel impacts the solute flux downstream and thereby induces spatial correlations in supply increase.

\begin{figure}[t]
\includegraphics[width=0.5\textwidth]{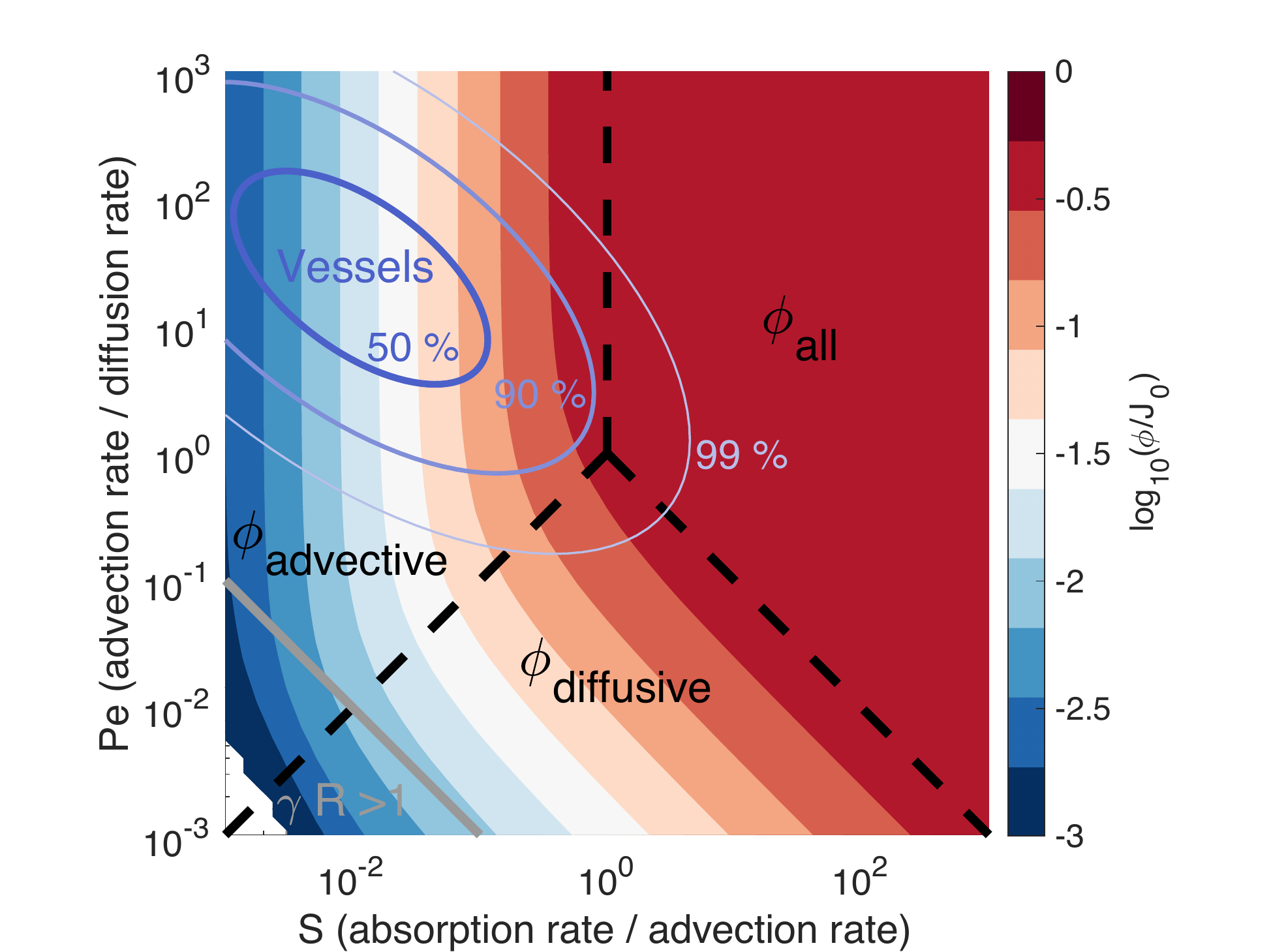}
\caption{ Supply $\phi$ by a single vessel can be partitioned into three distinctive regimes as a function of dimensionless parameters characterizing flow and absorption, $\Pe=\frac{\bar{U}L}{\kappa}$ and $S=\frac{\kappa \gamma L}{R \bar{U}}$. Dotted lines indicate separation of regimes. Remaining non-dimensional parameter fixed at $\alpha=0.001$. Error ellipsoids contain the annotated percentage of vessels of the here considered rat brain microvasculature \cite{blinder_cortical_2013} with physiological parameters for $\gamma$ and $\kappa$, see main text.}
\label{fig:vesselregime}
\end{figure}
To understand how a change in flow induces changes in solute flux and supply dynamics, we first focus on a single vessel. We assume that the flow is laminar with  longitudinal velocity profile  $U(r)=2\bar{U}(1-(r/R)^2)$ \cite{chang_optimal_2017,obrist_red_2010}, where $\bar{U}$ denotes the cross-sectional averaged longitudinal flow velocity. The dispersion of soluble molecules of concentration $C$ by the fluid flow within a tubular vessel of radius $R$ and length $L$ is then given by
\begin{align}
\frac{\partial C}{\partial t}+U(r)\frac{\partial C}{\partial z}=\kappa \nabla^2 C,
\label{eq:diffadv}
\end{align}
where $\kappa$ denotes the molecular diffusivity of the solute, and $r$ and $z$ parameterize the radial and longitudinal component of the vessel. The soluble molecule is absorbed at the vessel boundary, following 
\begin{align}
\kappa \left.\frac{\partial C}{\partial r}\right|_{r=R}+\kappa\gamma C(R)=0,
\label{eq:boundarycondition}
\end{align}
with absorption parameter $\gamma$. In analogy to the derivation of \textit{Taylor Dispersion} \cite{taylor_dispersion_1953,aris_dispersion_1956,meigel_flow_2018}, we simplify the multidimensional diffusion-advection for $C=\bar{C}+\tilde{C}$ to an equation for the cross-sectionally averaged concentration $\bar{C}$ if the cross-sectional variations of the concentration $\tilde{C}$ are much smaller than the averaged concentration itself. This is true if the time scale to diffuse radially within the vessel is much shorter than the time scale of advection along the vessel, ${R^2}/{\kappa}\ll {L}/{\bar{U}}$, if the vessel itself can be characterized as a long, slender vessel, $R \ll L$, and if the absorption parameter is small enough to keep a shallow gradient in concentration across the vessel's cross-section $\gamma R\ll 1$, which states that the length scale of absorption is much bigger than the vessel radius. All these approximations are valid for the rat brain microvasculature example considered here \cite{blinder_cortical_2013}. With these assumptions, the concentration profile along the vessel approaches a steady state over a timescale $L/\bar{U}$ given by (see the Supplemental Material~{S1} for derivation)
\begin{align}
\bar{C}(z)&=C_0\exp \left({- \beta(\Pe,S,\alpha)\frac{z}{L}   } \right), \label{eqn_steadystate} \\ 
\beta(\Pe,S,\alpha) &= \frac{24 \cdot \text{Pe} }{48  + \frac{\alpha^2}{S^2} } \left( \sqrt{1+ \frac{8 S}{\text{Pe}}+ \frac{\alpha^2}{6 \Pe S} } -1 \right), \label{eqn:exponentsteadystate}
\end{align}
where $\Pe={\bar{U}L}/{\kappa}$ is the
 P\'eclet number,  $\alpha = \gamma L$,  and \linebreak $S={\kappa \gamma L}/{R \bar{U}}$ measures the ratio of absorption rate to advection rate. Note, that the concentration decays along the vessel starting from an initial concentration $C_0$ that itself is determined by the solute flux entering a vessel $J_0$. 
Also for the solute influx into a vessel advective and diffusive transport contribute,
\begin{align}
    J_0=\pi R^2C_0\left(\bar{U}+\frac{\kappa \beta}{L}\right)=\pi R^2 C_0 \bar{U}\left(1+\frac{\beta}{\text{Pe}}\right) . \label{eqn:soluteinflux}
\end{align}
We define as supply of a vessel $\phi$ the integrated diffusive flux through the entire vessel surface $\mathcal{S}$ of the cylindrical vessel,
\begin{align}
   \phi = - \int_{\mathcal{S}} \kappa \left.\frac{\partial C}{\partial r}\right|_{r=R} ~2\pi R \text{d} z. \label{Eq:DefinePhi}
\end{align}
resulting in,
\begin{align}
\phi=& J_0\frac{1}{1+\frac{\beta}{\Pe}} \cdot \left( \frac{ \frac{\alpha^2}{12 S \Pe}  + 2 \frac{S}{\beta}}{1 +  \frac{\alpha^2}{4 S \Pe}} \right)
 \cdot   \left(1 - \exp \left( - \beta \right)  \right).
\label{eq:GrandAbsorb}
\end{align}

For physical intuition on how flow and vessel properties affect supply, we partition the phase space of supply dynamics spanned by $\Pe$ and $S$ into three regimes, keeping $\alpha$ fixed, see Fig.~{\ref{fig:vesselregime}}.
At large values of $S \gg 1$ and $S\gg 1/\Pe$ the solute decays very quickly along the vessel.
Here, all solute that flows into the vessel of cross-sectional area $\pi R^2$ is absorbed at the wall, here denoted {\sl all-absorbing regime}
\begin{align}
\phi_{\mathrm{all}} \approx  J_0 = \pi R^2 C_0 \bar{U}\left(1+\frac{\beta}{\text{Pe}}\right). \label{eq:Approx:Allabsorb} 
\end{align}
For a network this implies that after a vessel in this regime, no solute for further absorption downstream of this vessel is available, which indeed is physiologically rare, $1.0\%$ in the rat brain microvasculature considered here.
A second regime occurs at $\Pe\ll 1/S$, $\Pe\ll S$ where diffusive transport dominates, here denoted {\sl diffusive regime}. We distinguish a third regime, which we denote {\sl advective regime} where advective transport dominates, defined by $S\ll1$ and $S\ll \Pe$. In both cases the solute decay is very shallow, $\beta\ll 1$ in Eqs.~\eqref{eqn_steadystate}, {\eqref{Eq:DefinePhi}}, resulting in supply independent of flow velocity, except for the dependence on the initial concentration $C_0$ 
\begin{align}
    \phi_{\mathrm{advective}}\approx \phi_{\mathrm{diffusive}}\approx 2\pi R L \cdot \kappa \gamma \cdot C_0.  \label{eq:Approx:C0Absorption}
\end{align}
Yet, note that the reason for the solute decay, i.e. $\beta$ being small, arises from entirely different transport dynamics, see Fig.~\ref{fig:vesselregime}. This is reflected in the very different relation between initial solute concentration at the start of the vessel $C_0$ and the solute influx $J_0$ for the two regimes (see the Supplemental Material S1 for derivation)
\begin{align}
     &J_{0,\mathrm{advective}}\approx C_0 \cdot \pi R^2 \cdot \bar{U}, \label{eq:J0adv} \\
     &J_{0,\mathrm{diffusive}}\approx   C_0 \cdot \pi R^{\frac{3}{2}} \cdot \kappa \sqrt{2 \gamma}.\label{eq:J0diff} 
\end{align}
Hence, under constant solute influx $J_0$ the diffusive and the advective regime show a fundamentally different, yet both non-linear dependence on the vessel radius,
\begin{align}
&\phi_{\mathrm{advective}}\approx J_0 \frac{2 \gamma \kappa L}{ R \bar{U}}, \label{eq:Approx:Advection}\\
&\phi_{\mathrm{diffusive}}\approx J_0 \frac{\sqrt{2\gamma} L}{\sqrt{R}}.  \label{eq:Approx:Diffusion} 
\end{align}
Based on these results for a single vessel we expect largely varying increase in supply in response to vessel dilation. The coupling of flows and solute flux in a network is likely to make supply changes even more complex.

Within a network not only fluid flows are coupled with every network node obeying Kirchhoff's law $\sum_j \pi R_{\mathrm{in},j}^2 U_{\mathrm{in},j}=\sum_k \pi R_{\mathrm{out},k}^2 U_{\mathrm{out},k}$ but also solute flux $J$ is conserved at every node $\sum_j J_{\mathrm{in},j}=\sum_k J_{0,k}$. Here, the solute influx $J_{\mathrm{in},j}$ is determined by the inlet's vessel inflow $J_{0,j}$ upstream reduced by the amount of supply, $\phi_j$, via that vessel, see Eq.~\eqref{eq:GrandAbsorb}. The influxes $J_{0,k}$ downstream a node, defined by Eq.~\eqref{eqn:soluteinflux}, follow from the solute concentration at the network node $C_0$, given by
\begin{align}
C_0= \frac{\sum_j J_{\mathrm{in},j} }{\sum_k \pi R_{\mathrm{out},k}^2 (\bar{U}_{\mathrm{out},k}+\kappa \beta_{\mathrm{out},k}/L_{\mathrm{out},k})}.
\label{eq:initConc}
\end{align}
Thus, solute fluxes are subsequently propagated from network inlets throughout the network.

To now investigate the impact of single vessel dilation on supply within a network, we turn to an experimentally mapped rat brain microvasculature \cite{blinder_cortical_2013}. The data specifies $R$, $U$, and $L$  for all vessels as well as the pressures at network inlets and outlets. Focussing on glucose as primary demand, we account for glucose's diffusion constant $\kappa=\SI{6e-10}{\meter\squared\per\second}$ \cite{stein_channels_2012}
 and estimate glucose's permeability rate and include $\gamma = \SI{200}{\per\meter}$, see Supplemental Material S2.
 Interestingly, we find $98\%$ of all vessels to be in the advective regime. Is there a functional property that makes the advective regime stand out?

\begin{figure}[t]
\includegraphics[width=0.5\textwidth]{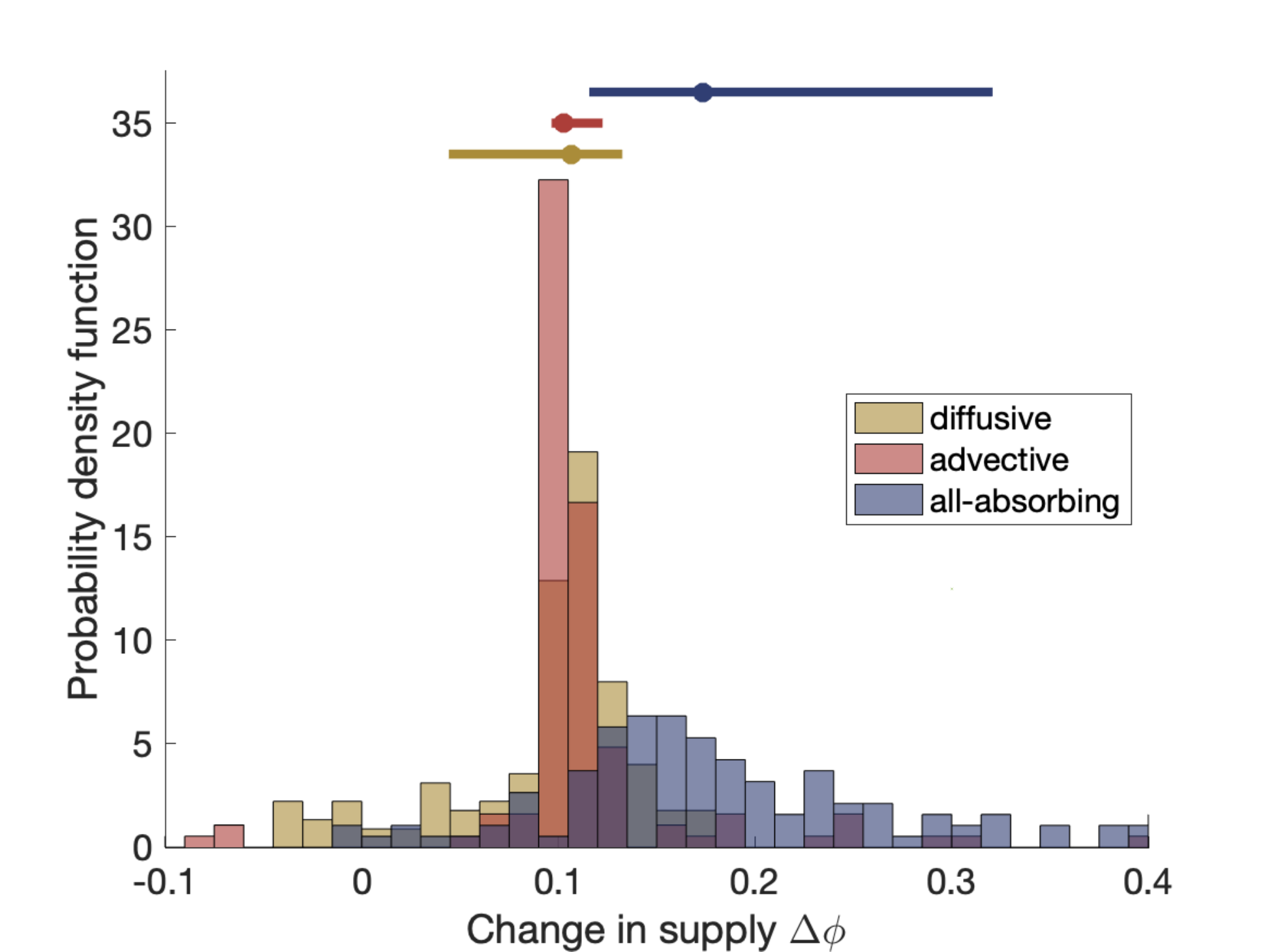}
\caption{ The advective regime is robust in increasing supply by dilation. Histogram of change in supply $\Delta \phi$ due to a single vessel dilating by $10\%$. Lines indicate a range covering $69\%$ with both a percentage of $15.5\%$ showing a lower or higher supply outside the indicated range.
Big dots indicate the median, with values of $0.17$, $0.10$, and $0.11$
 for the all-absorbing, advective, and diffusive regime, respectively. For each histogram 120
 vessels of the respective regime were randomly chosen and dilated.}
\label{fig:Histdilabs}
\end{figure}
We next quantify the change in supply due to vessel radius dilation in a capillary bed excerpt of the mapped rat brain microvasculature excluding pial and penetrating vessels. To this end, we use the pressures given in the data set \cite{blinder_cortical_2013} and impose the pressure values at inlet and outlet vessels of a network excerpt. To be consistent with the flows determined within the data set we use a modified hydraulic vessel resistance to account for additional blood hematocrit resistance \cite{pries_biophysical_1996,Rubenstein:2015} in accordance with Blinder et al.~\cite{blinder_cortical_2013}. Note, that a vessel's hydraulic resistance is only important to calculate fluid flow velocities within vessels but does not modify the supply dynamics derived above. Pressures and hydraulic resistances then fully determine the flow velocities throughout the network due to Kirchhoff's law.  

To identify differences in the behaviour of the three supply regimes that may justify the physiological abundance of the advective regime, we sample the effect of vessel dilation for all three regimes, drawing randomly 120 vessels in each regime out of the total number of 21793 vessels. The sheer total number of vessels allows us to sample underrepresented diffusive and all-absorbing regime without introducing a statistical bias due to sample size. Each vessels radius is dilated by 10\%, and the flow and solute flux is recalculated throughout the network keeping the networks inlet and outlet pressures fixed. The relative change in supply in the dilated vessel itself is evaluated in a histogram, see Fig.~{\ref{fig:Histdilabs}}. Vessels in the all-absorbing regime show a broad response to vessel dilation. Vessels in the advective regime, in contrast, peak sharply at a robust 10\% increase in supply, $\Delta \phi = 0.1$. The diffusive regime is also somewhat peaked around $\Delta \phi = 0.1$, but in addition shows a significant amount of vessels with smaller supply increase of $\Delta \phi <0.1$. Particularly the advective regime shows a robust increase in supply matching the increase in vessel diameter independent of the vessels' exact position within the network topology. This observation is robust against changes in the choice of the diffusion constant and permeability rate, see Supplemental Material~{S6}. 

Despite our expectations of a non-linear change in supply from single vessel dynamics, Eqs.~{\eqref{eq:Approx:Advection}}, {\eqref{eq:Approx:Diffusion}}, we find a robust increase of 10\% for 10\% vessel dilation, which would be reconciled within Eq.~{\eqref{eq:Approx:C0Absorption}}, if the initial concentration at the inlet of a dilating vessel $C_0$, Eq.~\eqref{eq:initConc}, stays constant despite changes in flow and solute flux throughout the network. Which network properties allow $C_0$ to stay constant? What makes the advective regime more robust than the diffusive?

Let us consider a network node, where all vessels are in the advective regime with one inlet vessel and two outlet vessels, out of the latter one is being dilated. Following Eq.~\eqref{eq:initConc} and the simplification of the solute fluxes from Eq.~\eqref{eq:J0adv} for the advective regime the initial concentration at the node is
\begin{align}
    C_0 \approx C_{\mathrm{in}} \frac{\pi R_{\mathrm{in}}^2 U_{\mathrm{in}}}{\sum_k \pi R_{\mathrm{out},k}^2 U_{\mathrm{out},k}} =C_{\mathrm{in}},
    \label{eq:C0adv}
\end{align}
where Kirchhoff's law was used for further simplification. Hence, even though vessel radius dilation induces changes in the flow, $C_0 \approx C_{\mathrm{in}}$ remains unchanged, though
$C_{\mathrm{in}}$ might be affected by upstream changes in the supply. However, we find that upstream effects on $C_{\mathrm{in}}$ are small if the upstream vessels are in the advective or diffusive regime, see Supplemental Material~{S3} and~{S5}, which leaves  $C_{\mathrm{in}}$ and thus $C_0$ approximately constant during vessel dilation.
This result generalizes to good approximation to the case where the non-dilating outlet vessel is in the diffusive rather than in the advective regime, see Supplemental Material~{S3}.
Note, that the case where two inlet vessels merge into one outlet vessels is fundamentally different, as then the initial concentration at the node is a mixture from the two inlet vessels. Dilation of the outlet vessel changes flow in inlets differently and thereby changes the mixing ratio non-linearly. Physiologically, we find this pattern especially closer toward venules. Taken together, these analytical results are in agreement with the statistics of Fig.~\ref{fig:Histdilabs} and explain in particular the robust increase in supply by dilation if the vessel is in the advective regime.

We next probe why the diffusive regime is less robust and revisit the setting of one inlet and one outlet in the advective regime, and the second outlet in the diffusive regime. But now we compute the initial concentration at the node given that we dilate the vessel in the diffusive regime,
\begin{align}
    C_0 \approx C_{\mathrm{in,adv}} \frac{\pi R_{\mathrm{in,adv}}^2 U_{\mathrm{in,adv}}}{ \pi R_{\mathrm{out,adv}}^2 U_{\mathrm{out,adv}}+  \pi R_{\mathrm{out,dif}}^{\frac{3}{2}} \kappa\sqrt{\gamma} }.
\end{align}
Now the dilation of the vessel in the diffusive regime increases the denominator and thus leads to a decrease in resulting $C_0$, rendering the diffusive vessel's response less robust compared to the advective. The same effect happens if all vessels at a node are in the diffusive regime, even more so as no vessel in the advective regime can buffer the dilation and diffusion dominated solute flux independent of flow velocity, see Eq.~\eqref{eq:J0diff}. 
Together, these analytical arguments explain why the diffusive regime yields a less robust increase in supply upon vessel dilation.

\begin{figure}
\includegraphics[width=0.45\textwidth]{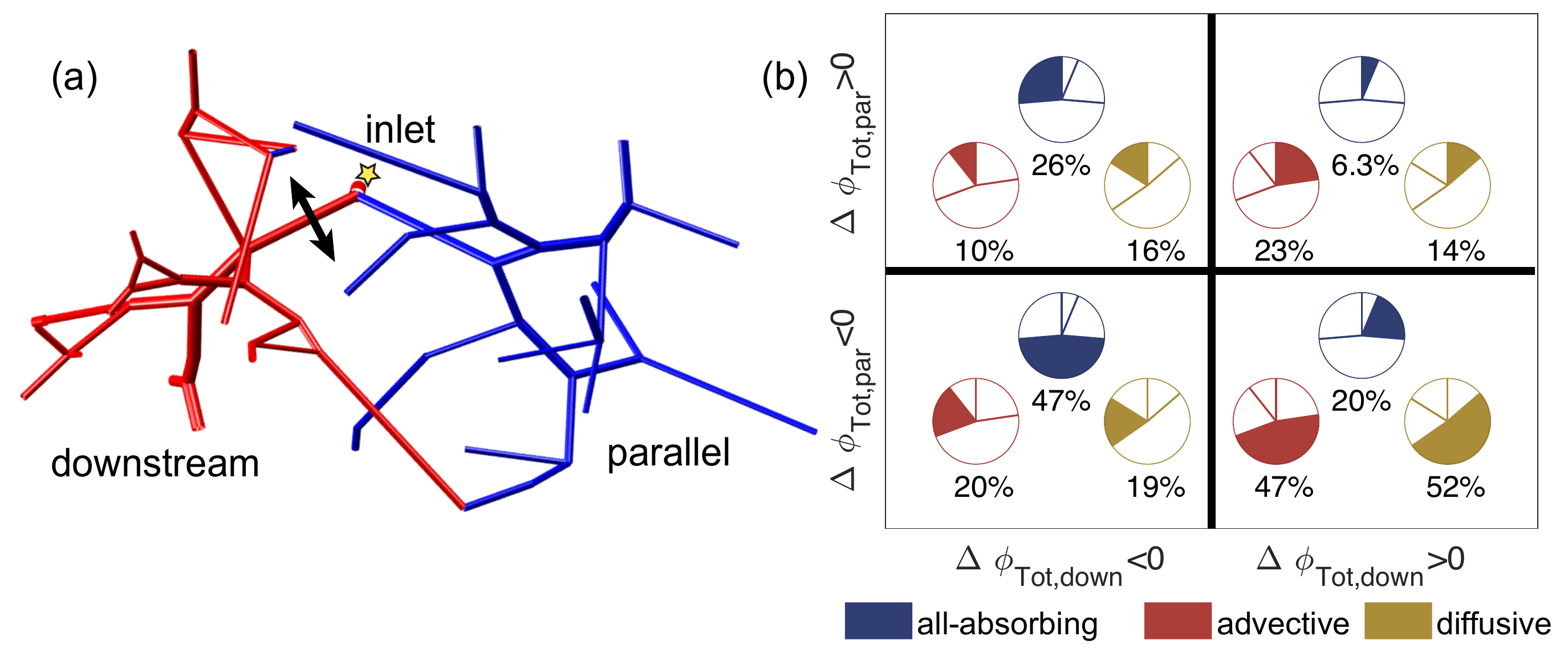}
\caption{ Advective and diffusive regime robustly increase supply downstream of a dilating vessel at the cost of decreasing supply in parallel vessels. (a) Enlargement of microvasculature excerpt exemplifying the neighborhood change in supply due to a single vessel dilation of $10\%$ (advective regime, black arrow). Inlet marked by yellow star. Blue denotes a decrease, red an increase in supply in the individual vessels. The total change in supply is $\Delta \phi_{\text{tot}}=6.4\%$ in the downstream vessels and $\phi_{\text{tot}}=0.8\%$ in the parallel vessels. Change in $C_0$ for the dilating vessel is below $\Delta C_0 < 3\times 10^{-4}$. (b) Neighborhood statistics of supply increase `+' or decrease `-' due to a dilating vessel in the respective regime. Evaluated is the overall change in supply in up to four vessels downstream or parallel to the dilated vessel chosen at the main inlet of a loop, respectively. The dilated vessel itself is excluded from the statistics here.}
\label{fig:Upstream}
\end{figure}

We found in Fig.~{\ref{fig:Histdilabs}}, that the supply in upstream vessel remains approximately constant during a single vessel dilation. What is the effect on vessels downsteam the dilated vessel? For this, we focus on the dilating vessel's immediate neighborhood and find that change in supply is spatially correlated, Fig.~{\ref{fig:Upstream}}. We distinguish the vessels in the direct neighbourhood of the dilated vessel in two categories: \textit{downstream vessels} are vessels that are located directly downstream of the dilated vessel and \textit{parallel vessels} are vessels that are downstream the node the dilated vessels branches off, but not downstream the dilated vessel itself. The microvasculature data set is known to show predominantly loop topologies, with a median size of eight vessels within a loop \cite{blinder_cortical_2013}. We thus considered only vessels with a topological distance of four vessels to the dilated vessel for the analysis of the immediate neighbourhood. We find that the typical response of a dilating vessel in both advective and diffusive regime is to increase supply downstream at the cost of reducing supply in the parallel vessels, Fig.~{\ref{fig:Upstream}}~(b). 
More solute is drawn along the branch of the loop containing a dilating vessel than the dilating vessel itself is taking up, which increases the supply in downstream vessels. This is at the expense of the vessels in the parallel branch, reducing the supply there. See also Supplemental Material~{S4}.
While this applies qualitatively, the strength of this effect depends on the exact network topology.

We here provided a theoretical framework to investigate supply dynamics in a dynamically adapting tubular network, where flows are globally coupled by topology. We find that individual vessels can be classified in three regimes by vessel geometry and flow rate. Among those particularly the regime governed by advective transport - and to lesser extend also the regime governed by diffusive transport - yield a robust increase in supply upon vessel dilation within the dilating vessel, notably leaving the supply pattern upstream unchanged and increasing supply immediately downstream. Interestingly, the most robust advective regime is found to dominate in brain microvasculature.  Our findings therefore promote that vessel dilation results in a robust increase in supply independent of the exact position of the vessel in the network. Our results are important for understanding the link between neural activity and patterns of change in supply invoked by vessel dilations and changes in blood flow underlying fMRI. Moreover, our framework is instrumental to predict drug delivery, design blood vessel architecture in synthetic organs but may also open entire new avenues for the programming of soft robotics and smart materials.
\begin{acknowledgments}
We thank David Kleinfeld and collaborators for sharing their data on the rat brain microvasculature with us. 
This work was supported by the Max Planck Society and the National Science Foundation Division of Mathematical Sciences DMS 1411694 and DMS 1715477. K.A. further acknowledges the stimulating environment of American Institute of Mathematics' Square Meetings. M.P.B. is an investigator of the Simons Foundation. 
\end{acknowledgments}


\begin{thebibliography}{33}%
\makeatletter
\providecommand \@ifxundefined [1]{%
 \@ifx{#1\undefined}
}%
\providecommand \@ifnum [1]{%
 \ifnum #1\expandafter \@firstoftwo
 \else \expandafter \@secondoftwo
 \fi
}%
\providecommand \@ifx [1]{%
 \ifx #1\expandafter \@firstoftwo
 \else \expandafter \@secondoftwo
 \fi
}%
\providecommand \natexlab [1]{#1}%
\providecommand \enquote  [1]{``#1''}%
\providecommand \bibnamefont  [1]{#1}%
\providecommand \bibfnamefont [1]{#1}%
\providecommand \citenamefont [1]{#1}%
\providecommand \href@noop [0]{\@secondoftwo}%
\providecommand \href [0]{\begingroup \@sanitize@url \@href}%
\providecommand \@href[1]{\@@startlink{#1}\@@href}%
\providecommand \@@href[1]{\endgroup#1\@@endlink}%
\providecommand \@sanitize@url [0]{\catcode `\\12\catcode `\$12\catcode
  `\&12\catcode `\#12\catcode `\^12\catcode `\_12\catcode `\%12\relax}%
\providecommand \@@startlink[1]{}%
\providecommand \@@endlink[0]{}%
\providecommand \url  [0]{\begingroup\@sanitize@url \@url }%
\providecommand \@url [1]{\endgroup\@href {#1}{\urlprefix }}%
\providecommand \urlprefix  [0]{URL }%
\providecommand \Eprint [0]{\href }%
\providecommand \doibase [0]{http://dx.doi.org/}%
\providecommand \selectlanguage [0]{\@gobble}%
\providecommand \bibinfo  [0]{\@secondoftwo}%
\providecommand \bibfield  [0]{\@secondoftwo}%
\providecommand \translation [1]{[#1]}%
\providecommand \BibitemOpen [0]{}%
\providecommand \bibitemStop [0]{}%
\providecommand \bibitemNoStop [0]{.\EOS\space}%
\providecommand \EOS [0]{\spacefactor3000\relax}%
\providecommand \BibitemShut  [1]{\csname bibitem#1\endcsname}%
\let\auto@bib@innerbib\@empty
\bibitem [{\citenamefont {Murray}(1926)}]{Murray:1926tj}%
  \BibitemOpen
  \bibfield  {author} {\bibinfo {author} {\bibfnamefont {C.~D.}\ \bibnamefont
  {Murray}},\ }\href@noop {} {\bibfield  {journal} {\bibinfo  {journal} {Proc.
  Natl. Acad. Sci. U.S.A.}\ }\textbf {\bibinfo {volume} {12}},\ \bibinfo
  {pages} {207} (\bibinfo {year} {1926})}\BibitemShut {NoStop}%
\bibitem [{\citenamefont {West}\ \emph {et~al.}(1997)\citenamefont {West},
  \citenamefont {Brown},\ and\ \citenamefont {Enquist}}]{West:1997ww}%
  \BibitemOpen
  \bibfield  {author} {\bibinfo {author} {\bibfnamefont {G.~B.}\ \bibnamefont
  {West}}, \bibinfo {author} {\bibfnamefont {J.~H.}\ \bibnamefont {Brown}}, \
  and\ \bibinfo {author} {\bibfnamefont {B.~J.}\ \bibnamefont {Enquist}},\
  }\href@noop {} {\bibfield  {journal} {\bibinfo  {journal} {Science}\ }\textbf
  {\bibinfo {volume} {276}},\ \bibinfo {pages} {122} (\bibinfo {year}
  {1997})}\BibitemShut {NoStop}%
\bibitem [{\citenamefont {Hu}\ and\ \citenamefont {Cai}(2013)}]{Hu:2013io}%
  \BibitemOpen
  \bibfield  {author} {\bibinfo {author} {\bibfnamefont {D.}~\bibnamefont
  {Hu}}\ and\ \bibinfo {author} {\bibfnamefont {D.}~\bibnamefont {Cai}},\
  }\href@noop {} {\bibfield  {journal} {\bibinfo  {journal} {Phys. Rev. Lett.}\
  }\textbf {\bibinfo {volume} {111}},\ \bibinfo {pages} {138701} (\bibinfo
  {year} {2013})}\BibitemShut {NoStop}%
\bibitem [{\citenamefont {Ronellenfitsch}\ and\ \citenamefont
  {Katifori}(2016)}]{Ronellenfitsch:2016hh}%
  \BibitemOpen
  \bibfield  {author} {\bibinfo {author} {\bibfnamefont {H.}~\bibnamefont
  {Ronellenfitsch}}\ and\ \bibinfo {author} {\bibfnamefont {E.}~\bibnamefont
  {Katifori}},\ }\href@noop {} {\bibfield  {journal} {\bibinfo  {journal}
  {Phys. Rev. Lett.}\ }\textbf {\bibinfo {volume} {117}},\ \bibinfo {pages}
  {138301} (\bibinfo {year} {2016})}\BibitemShut {NoStop}%
\bibitem [{\citenamefont {Zheng}\ \emph {et~al.}(2017)\citenamefont {Zheng},
  \citenamefont {Shen}, \citenamefont {Wang}, \citenamefont {Dunphy},
  \citenamefont {Hasan}, \citenamefont {Brinker}, \citenamefont {Li},\ and\
  \citenamefont {Su}}]{Zheng:2017jg}%
  \BibitemOpen
  \bibfield  {author} {\bibinfo {author} {\bibfnamefont {X.}~\bibnamefont
  {Zheng}}, \bibinfo {author} {\bibfnamefont {G.}~\bibnamefont {Shen}},
  \bibinfo {author} {\bibfnamefont {C.}~\bibnamefont {Wang}}, \bibinfo {author}
  {\bibfnamefont {D.}~\bibnamefont {Dunphy}}, \bibinfo {author} {\bibfnamefont
  {T.}~\bibnamefont {Hasan}}, \bibinfo {author} {\bibfnamefont {C.~J.}\
  \bibnamefont {Brinker}}, \bibinfo {author} {\bibfnamefont {Y.}~\bibnamefont
  {Li}}, \ and\ \bibinfo {author} {\bibfnamefont {B.-L.}\ \bibnamefont {Su}},\
  }\href@noop {} {\bibfield  {journal} {\bibinfo  {journal} {Nat. Commun.}\
  }\textbf {\bibinfo {volume} {8}},\ \bibinfo {pages} {1} (\bibinfo {year}
  {2017})}\BibitemShut {NoStop}%
\bibitem [{\citenamefont {Boynton}\ \emph {et~al.}(1996)\citenamefont
  {Boynton}, \citenamefont {Engel}, \citenamefont {Glover},\ and\ \citenamefont
  {Heeger}}]{boynton_linear_1996}%
  \BibitemOpen
  \bibfield  {author} {\bibinfo {author} {\bibfnamefont {G.~M.}\ \bibnamefont
  {Boynton}}, \bibinfo {author} {\bibfnamefont {S.~A.}\ \bibnamefont {Engel}},
  \bibinfo {author} {\bibfnamefont {G.~H.}\ \bibnamefont {Glover}}, \ and\
  \bibinfo {author} {\bibfnamefont {D.~J.}\ \bibnamefont {Heeger}},\ }\href
  {\doibase 10.1523/JNEUROSCI.16-13-04207.1996} {\bibfield  {journal} {\bibinfo
   {journal} {J. Neurosci.}\ }\textbf {\bibinfo {volume} {16}},\ \bibinfo
  {pages} {4207} (\bibinfo {year} {1996})}\BibitemShut {NoStop}%
\bibitem [{\citenamefont {Raichle}\ and\ \citenamefont
  {Mintun}(2006)}]{raichle_brain_2006}%
  \BibitemOpen
  \bibfield  {author} {\bibinfo {author} {\bibfnamefont {M.~E.}\ \bibnamefont
  {Raichle}}\ and\ \bibinfo {author} {\bibfnamefont {M.~A.}\ \bibnamefont
  {Mintun}},\ }\href {\doibase 10.1146/annurev.neuro.29.051605.112819}
  {\bibfield  {journal} {\bibinfo  {journal} {Annu. Rev. Neurosci.}\ }\textbf
  {\bibinfo {volume} {29}},\ \bibinfo {pages} {449} (\bibinfo {year}
  {2006})}\BibitemShut {NoStop}%
\bibitem [{\citenamefont {Blinder}\ \emph {et~al.}(2013)\citenamefont
  {Blinder}, \citenamefont {Tsai}, \citenamefont {Kaufhold}, \citenamefont
  {Knutsen}, \citenamefont {Suhl},\ and\ \citenamefont
  {Kleinfeld}}]{blinder_cortical_2013}%
  \BibitemOpen
  \bibfield  {author} {\bibinfo {author} {\bibfnamefont {P.}~\bibnamefont
  {Blinder}}, \bibinfo {author} {\bibfnamefont {P.~S.}\ \bibnamefont {Tsai}},
  \bibinfo {author} {\bibfnamefont {J.~P.}\ \bibnamefont {Kaufhold}}, \bibinfo
  {author} {\bibfnamefont {P.~M.}\ \bibnamefont {Knutsen}}, \bibinfo {author}
  {\bibfnamefont {H.}~\bibnamefont {Suhl}}, \ and\ \bibinfo {author}
  {\bibfnamefont {D.}~\bibnamefont {Kleinfeld}},\ }\href {\doibase
  10.1038/nn.3426} {\bibfield  {journal} {\bibinfo  {journal} {Nat. Neurosci.}\
  }\textbf {\bibinfo {volume} {16}},\ \bibinfo {pages} {889} (\bibinfo {year}
  {2013})}\BibitemShut {NoStop}%
\bibitem [{\citenamefont {Logothetis}(2008)}]{logothetis_what_2008}%
  \BibitemOpen
  \bibfield  {author} {\bibinfo {author} {\bibfnamefont {N.~K.}\ \bibnamefont
  {Logothetis}},\ }\href {\doibase 10.1038/nature06976} {\bibfield  {journal}
  {\bibinfo  {journal} {Nature}\ }\textbf {\bibinfo {volume} {453}},\ \bibinfo
  {pages} {869} (\bibinfo {year} {2008})}\BibitemShut {NoStop}%
\bibitem [{\citenamefont {MacVicar}\ and\ \citenamefont
  {Newman}(2015)}]{macvicar_astrocyte_2015}%
  \BibitemOpen
  \bibfield  {author} {\bibinfo {author} {\bibfnamefont {B.~A.}\ \bibnamefont
  {MacVicar}}\ and\ \bibinfo {author} {\bibfnamefont {E.~A.}\ \bibnamefont
  {Newman}},\ }\href {\doibase 10.1101/cshperspect.a020388} {\bibfield
  {journal} {\bibinfo  {journal} {Cold Spring Harb. Perspect. Biol.}\ }\textbf
  {\bibinfo {volume} {7}},\ \bibinfo {pages} {a020388} (\bibinfo {year}
  {2015})}\BibitemShut {NoStop}%
\bibitem [{\citenamefont {Cai}\ \emph {et~al.}(2018)\citenamefont {Cai},
  \citenamefont {Fordsmann}, \citenamefont {Jensen}, \citenamefont {Gesslein},
  \citenamefont {Lønstrup}, \citenamefont {Hald}, \citenamefont {Zambach},
  \citenamefont {Brodin},\ and\ \citenamefont
  {Lauritzen}}]{cai_stimulation-induced_2018}%
  \BibitemOpen
  \bibfield  {author} {\bibinfo {author} {\bibfnamefont {C.}~\bibnamefont
  {Cai}}, \bibinfo {author} {\bibfnamefont {J.~C.}\ \bibnamefont {Fordsmann}},
  \bibinfo {author} {\bibfnamefont {S.~H.}\ \bibnamefont {Jensen}}, \bibinfo
  {author} {\bibfnamefont {B.}~\bibnamefont {Gesslein}}, \bibinfo {author}
  {\bibfnamefont {M.}~\bibnamefont {Lønstrup}}, \bibinfo {author}
  {\bibfnamefont {B.~O.}\ \bibnamefont {Hald}}, \bibinfo {author}
  {\bibfnamefont {S.~A.}\ \bibnamefont {Zambach}}, \bibinfo {author}
  {\bibfnamefont {B.}~\bibnamefont {Brodin}}, \ and\ \bibinfo {author}
  {\bibfnamefont {M.~J.}\ \bibnamefont {Lauritzen}},\ }\href {\doibase
  10.1073/pnas.1707702115} {\bibfield  {journal} {\bibinfo  {journal} {Proc.
  Natl. Acad. Sci. U.S.A.}\
  }\textbf {\bibinfo {volume} {115}},\ \bibinfo {pages} {E5796} (\bibinfo
  {year} {2018})}\BibitemShut {NoStop}%
\bibitem [{\citenamefont {Hill}\ \emph {et~al.}(2015)\citenamefont {Hill},
  \citenamefont {Tong}, \citenamefont {Yuan}, \citenamefont {Murikinati},
  \citenamefont {Gupta},\ and\ \citenamefont
  {Grutzendler}}]{hill_regional_2015}%
  \BibitemOpen
  \bibfield  {author} {\bibinfo {author} {\bibfnamefont {R.~A.}\ \bibnamefont
  {Hill}}, \bibinfo {author} {\bibfnamefont {L.}~\bibnamefont {Tong}}, \bibinfo
  {author} {\bibfnamefont {P.}~\bibnamefont {Yuan}}, \bibinfo {author}
  {\bibfnamefont {S.}~\bibnamefont {Murikinati}}, \bibinfo {author}
  {\bibfnamefont {S.}~\bibnamefont {Gupta}}, \ and\ \bibinfo {author}
  {\bibfnamefont {J.}~\bibnamefont {Grutzendler}},\ }\href {\doibase
  10.1016/j.neuron.2015.06.001} {\bibfield  {journal} {\bibinfo  {journal}
  {Neuron}\ }\textbf {\bibinfo {volume} {87}},\ \bibinfo {pages} {95} (\bibinfo
  {year} {2015})}\BibitemShut {NoStop}%
\bibitem [{\citenamefont {Vanzetta}\ \emph {et~al.}(2005)\citenamefont
  {Vanzetta}, \citenamefont {Hildesheim},\ and\ \citenamefont
  {Grinvald}}]{vanzetta_compartment-resolved_2005}%
  \BibitemOpen
  \bibfield  {author} {\bibinfo {author} {\bibfnamefont {I.}~\bibnamefont
  {Vanzetta}}, \bibinfo {author} {\bibfnamefont {R.}~\bibnamefont
  {Hildesheim}}, \ and\ \bibinfo {author} {\bibfnamefont {A.}~\bibnamefont
  {Grinvald}},\ }\href {\doibase 10.1523/JNEUROSCI.3032-04.2005} {\bibfield
  {journal} {\bibinfo  {journal} {J. Neurosci.}\ }\textbf {\bibinfo {volume}
  {25}},\ \bibinfo {pages} {2233} (\bibinfo {year} {2005})}\BibitemShut
  {NoStop}%
\bibitem [{\citenamefont {Krogh}(1919)}]{krogh_number_1919}%
  \BibitemOpen
  \bibfield  {author} {\bibinfo {author} {\bibfnamefont {A.}~\bibnamefont
  {Krogh}},\ }\href {\doibase 10.1113/jphysiol.1919.sp001839} {\bibfield
  {journal} {\bibinfo  {journal} {J. Physiol.}\ }\textbf {\bibinfo {volume}
  {52}},\ \bibinfo {pages} {409} (\bibinfo {year} {1919})}\BibitemShut
  {NoStop}%
\bibitem [{\citenamefont {Schneider}\ \emph {et~al.}(2012)\citenamefont
  {Schneider}, \citenamefont {Reichold}, \citenamefont {Weber}, \citenamefont
  {Székely},\ and\ \citenamefont {Hirsch}}]{schneider_tissue_2012}%
  \BibitemOpen
  \bibfield  {author} {\bibinfo {author} {\bibfnamefont {M.}~\bibnamefont
  {Schneider}}, \bibinfo {author} {\bibfnamefont {J.}~\bibnamefont {Reichold}},
  \bibinfo {author} {\bibfnamefont {B.}~\bibnamefont {Weber}}, \bibinfo
  {author} {\bibfnamefont {G.}~\bibnamefont {Székely}}, \ and\ \bibinfo
  {author} {\bibfnamefont {S.}~\bibnamefont {Hirsch}},\ }\href {\doibase
  10.1016/j.media.2012.04.009} {\bibfield  {journal} {\bibinfo  {journal} {Med.
  Image Anal.}\ }\textbf {\bibinfo {volume} {16}},\ \bibinfo {pages} {1397}
  (\bibinfo {year} {2012})}\BibitemShut {NoStop}%
\bibitem [{\citenamefont {Taylor}(1953)}]{taylor_dispersion_1953}%
  \BibitemOpen
  \bibfield  {author} {\bibinfo {author} {\bibfnamefont {G.~I.}\ \bibnamefont
  {Taylor}},\ }\href {\doibase 10.1098/rspa.1953.0139} {\bibfield  {journal}
  {\bibinfo  {journal} {Proc. R. Soc. Lond. A}\ }\textbf {\bibinfo {volume}
  {219}},\ \bibinfo {pages} {186} (\bibinfo {year} {1953})}\BibitemShut
  {NoStop}%
\bibitem [{\citenamefont {Aris}(1956)}]{aris_dispersion_1956}%
  \BibitemOpen
  \bibfield  {author} {\bibinfo {author} {\bibfnamefont {R.}~\bibnamefont
  {Aris}},\ }\href {\doibase 10.1098/rspa.1956.0065} {\bibfield  {journal}
  {\bibinfo  {journal} {Proc. R. Soc. Lond. A}\ }\textbf {\bibinfo {volume}
  {235}},\ \bibinfo {pages} {67} (\bibinfo {year} {1956})}\BibitemShut
  {NoStop}%
\bibitem [{\citenamefont {Lungu}\ and\ \citenamefont
  {Moffatt}(1982)}]{e._m._lungu_effect_1982}%
  \BibitemOpen
  \bibfield  {author} {\bibinfo {author} {\bibfnamefont {E.~M.}\ \bibnamefont
  {Lungu}}\ and\ \bibinfo {author} {\bibfnamefont {H.~K.}\ \bibnamefont
  {Moffatt}},\ }\href {\doibase 10.1007/BF00042550} {\bibfield  {journal}
  {\bibinfo  {journal} {J. Eng. Math.}\ }\textbf {\bibinfo {volume} {16}},\
  \bibinfo {pages} {121} (\bibinfo {year} {1982})}\BibitemShut {NoStop}%
\bibitem [{\citenamefont {Shapiro}\ and\ \citenamefont
  {Brenner}(1986)}]{shapiro_taylor_1986}%
  \BibitemOpen
  \bibfield  {author} {\bibinfo {author} {\bibfnamefont {M.}~\bibnamefont
  {Shapiro}}\ and\ \bibinfo {author} {\bibfnamefont {H.}~\bibnamefont
  {Brenner}},\ }\href {\doibase 10.1016/0009-2509(86)85228-9} {\bibfield
  {journal} {\bibinfo  {journal} {Chem. Eng. Sci.}\ }\textbf {\bibinfo {volume}
  {41}},\ \bibinfo {pages} {1417} (\bibinfo {year} {1986})}\BibitemShut
  {NoStop}%
\bibitem [{\citenamefont {Mercer}\ and\ \citenamefont
  {Roberts}(1994)}]{mercer_complete_1994}%
  \BibitemOpen
  \bibfield  {author} {\bibinfo {author} {\bibfnamefont {G.~N.}\ \bibnamefont
  {Mercer}}\ and\ \bibinfo {author} {\bibfnamefont {A.~J.}\ \bibnamefont
  {Roberts}},\ }\href {\doibase 10.1007/BF03167234} {\bibfield  {journal}
  {\bibinfo  {journal} {Japan J. Indust. Appl. Math.}\ }\textbf {\bibinfo
  {volume} {11}},\ \bibinfo {pages} {499} (\bibinfo {year} {1994})}\BibitemShut
  {NoStop}%
\bibitem [{\citenamefont {Marbach}\ \emph {et~al.}(2016)\citenamefont
  {Marbach}, \citenamefont {Alim}, \citenamefont {Andrew}, \citenamefont
  {Pringle},\ and\ \citenamefont {Brenner}}]{marbach_pruning_2016}%
  \BibitemOpen
  \bibfield  {author} {\bibinfo {author} {\bibfnamefont {S.}~\bibnamefont
  {Marbach}}, \bibinfo {author} {\bibfnamefont {K.}~\bibnamefont {Alim}},
  \bibinfo {author} {\bibfnamefont {N.}~\bibnamefont {Andrew}}, \bibinfo
  {author} {\bibfnamefont {A.}~\bibnamefont {Pringle}}, \ and\ \bibinfo
  {author} {\bibfnamefont {M.~P.}\ \bibnamefont {Brenner}},\ }\href {\doibase
  10.1103/PhysRevLett.117.178103} {\bibfield  {journal} {\bibinfo  {journal}
  {Phys. Rev. Lett.}\ }\textbf {\bibinfo {volume} {117}},\ \bibinfo {pages}
  {178103} (\bibinfo {year} {2016})}\BibitemShut {NoStop}%
\bibitem [{\citenamefont {Fang}\ \emph {et~al.}(2008)\citenamefont {Fang},
  \citenamefont {Sakadzić}, \citenamefont {Ruvinskaya}, \citenamefont {Devor},
  \citenamefont {Dale},\ and\ \citenamefont {Boas}}]{fang_oxygen_2008}%
  \BibitemOpen
  \bibfield  {author} {\bibinfo {author} {\bibfnamefont {Q.}~\bibnamefont
  {Fang}}, \bibinfo {author} {\bibfnamefont {S.}~\bibnamefont {Sakadzić}},
  \bibinfo {author} {\bibfnamefont {L.}~\bibnamefont {Ruvinskaya}}, \bibinfo
  {author} {\bibfnamefont {A.}~\bibnamefont {Devor}}, \bibinfo {author}
  {\bibfnamefont {A.~M.}\ \bibnamefont {Dale}}, \ and\ \bibinfo {author}
  {\bibfnamefont {D.~A.}\ \bibnamefont {Boas}},\ }\href@noop {} {\bibfield
  {journal} {\bibinfo  {journal} {Opt. Express}\ }\textbf {\bibinfo {volume}
  {16}},\ \bibinfo {pages} {17530} (\bibinfo {year} {2008})}\BibitemShut
  {NoStop}%
\bibitem [{\citenamefont {Meigel}\ and\ \citenamefont
  {Alim}(2018)}]{meigel_flow_2018}%
  \BibitemOpen
  \bibfield  {author} {\bibinfo {author} {\bibfnamefont {F.~J.}\ \bibnamefont
  {Meigel}}\ and\ \bibinfo {author} {\bibfnamefont {K.}~\bibnamefont {Alim}},\
  }\href {\doibase 10.1098/rsif.2018.0075} {\bibfield  {journal} {\bibinfo
  {journal} {Roy. Soc. Interface}\ }\textbf {\bibinfo {volume} {15}},\ \bibinfo
  {pages} {20180075} (\bibinfo {year} {2018})}\BibitemShut {NoStop}%
\bibitem [{\citenamefont {He}\ \emph {et~al.}(2018)\citenamefont {He},
  \citenamefont {Wang}, \citenamefont {Chen}, \citenamefont {Pohmann},
  \citenamefont {Polimeni}, \citenamefont {Scheffler}, \citenamefont {Rosen},
  \citenamefont {Kleinfeld},\ and\ \citenamefont {Yu}}]{he_ultra-slow_2018}%
  \BibitemOpen
  \bibfield  {author} {\bibinfo {author} {\bibfnamefont {Y.}~\bibnamefont
  {He}}, \bibinfo {author} {\bibfnamefont {M.}~\bibnamefont {Wang}}, \bibinfo
  {author} {\bibfnamefont {X.}~\bibnamefont {Chen}}, \bibinfo {author}
  {\bibfnamefont {R.}~\bibnamefont {Pohmann}}, \bibinfo {author} {\bibfnamefont
  {J.~R.}\ \bibnamefont {Polimeni}}, \bibinfo {author} {\bibfnamefont
  {K.}~\bibnamefont {Scheffler}}, \bibinfo {author} {\bibfnamefont {B.~R.}\
  \bibnamefont {Rosen}}, \bibinfo {author} {\bibfnamefont {D.}~\bibnamefont
  {Kleinfeld}}, \ and\ \bibinfo {author} {\bibfnamefont {X.}~\bibnamefont
  {Yu}},\ }\href {\doibase 10.1016/j.neuron.2018.01.025} {\bibfield  {journal}
  {\bibinfo  {journal} {Neuron}\ }\textbf {\bibinfo {volume} {97}},\ \bibinfo
  {pages} {925} (\bibinfo {year} {2018})}\BibitemShut {NoStop}%
\bibitem [{\citenamefont {Tian}\ \emph {et~al.}(2010)\citenamefont {Tian},
  \citenamefont {Teng}, \citenamefont {May}, \citenamefont {Kurz},
  \citenamefont {Lu}, \citenamefont {Scadeng}, \citenamefont {Hillman},
  \citenamefont {Crespigny}, \citenamefont {D’Arceuil}, \citenamefont
  {Mandeville}, \citenamefont {Marota}, \citenamefont {Rosen}, \citenamefont
  {Liu}, \citenamefont {Boas}, \citenamefont {Buxton}, \citenamefont {Dale},\
  and\ \citenamefont {Devor}}]{tian_cortical_2010}%
  \BibitemOpen
  \bibfield  {author} {\bibinfo {author} {\bibfnamefont {P.}~\bibnamefont
  {Tian}}, \bibinfo {author} {\bibfnamefont {I.~C.}\ \bibnamefont {Teng}},
  \bibinfo {author} {\bibfnamefont {L.~D.}\ \bibnamefont {May}}, \bibinfo
  {author} {\bibfnamefont {R.}~\bibnamefont {Kurz}}, \bibinfo {author}
  {\bibfnamefont {K.}~\bibnamefont {Lu}}, \bibinfo {author} {\bibfnamefont
  {M.}~\bibnamefont {Scadeng}}, \bibinfo {author} {\bibfnamefont {E.~M.~C.}\
  \bibnamefont {Hillman}}, \bibinfo {author} {\bibfnamefont {A.~J.~D.}\
  \bibnamefont {Crespigny}}, \bibinfo {author} {\bibfnamefont {H.~E.}\
  \bibnamefont {D’Arceuil}}, \bibinfo {author} {\bibfnamefont {J.~B.}\
  \bibnamefont {Mandeville}}, \bibinfo {author} {\bibfnamefont {J.~J.~A.}\
  \bibnamefont {Marota}}, \bibinfo {author} {\bibfnamefont {B.~R.}\
  \bibnamefont {Rosen}}, \bibinfo {author} {\bibfnamefont {T.~T.}\ \bibnamefont
  {Liu}}, \bibinfo {author} {\bibfnamefont {D.~A.}\ \bibnamefont {Boas}},
  \bibinfo {author} {\bibfnamefont {R.~B.}\ \bibnamefont {Buxton}}, \bibinfo
  {author} {\bibfnamefont {A.~M.}\ \bibnamefont {Dale}}, \ and\ \bibinfo
  {author} {\bibfnamefont {A.}~\bibnamefont {Devor}},\ }\href {\doibase
  10.1073/pnas.1006735107} {\bibfield  {journal} {\bibinfo  {journal} {Proc.
  Natl. Acad. Sci. U.S.A.}\
  }\textbf {\bibinfo {volume} {107}},\ \bibinfo {pages} {15246} (\bibinfo
  {year} {2010})}\BibitemShut {NoStop}%
\bibitem [{\citenamefont {Paulson}\ \emph {et~al.}(2010)\citenamefont
  {Paulson}, \citenamefont {Hasselbalch}, \citenamefont {Rostrup},
  \citenamefont {Knudsen},\ and\ \citenamefont {Pelligrino}}]{Paulson}%
  \BibitemOpen
  \bibfield  {author} {\bibinfo {author} {\bibfnamefont {O.~B.}\ \bibnamefont
  {Paulson}}, \bibinfo {author} {\bibfnamefont {S.~G.}\ \bibnamefont
  {Hasselbalch}}, \bibinfo {author} {\bibfnamefont {E.}~\bibnamefont
  {Rostrup}}, \bibinfo {author} {\bibfnamefont {G.~M.}\ \bibnamefont
  {Knudsen}}, \ and\ \bibinfo {author} {\bibfnamefont {D.}~\bibnamefont
  {Pelligrino}},\ }\href@noop {} {\bibfield  {journal} {\bibinfo  {journal}
  {Journal of Cerebral Blood Flow \& Metabolism}\ }\textbf {\bibinfo {volume}
  {30}},\ \bibinfo {pages} {2} (\bibinfo {year} {2010})}\BibitemShut {NoStop}%
\bibitem [{\citenamefont {O’Herron}\ \emph {et~al.}(2016)\citenamefont
  {O’Herron}, \citenamefont {Chhatbar}, \citenamefont {Levy}, \citenamefont
  {Shen}, \citenamefont {Schramm}, \citenamefont {Lu},\ and\ \citenamefont
  {Kara}}]{oherron_neural_2016}%
  \BibitemOpen
  \bibfield  {author} {\bibinfo {author} {\bibfnamefont {P.}~\bibnamefont
  {O'Herron}}, \bibinfo {author} {\bibfnamefont {P.~Y.}\ \bibnamefont
  {Chhatbar}}, \bibinfo {author} {\bibfnamefont {M.}~\bibnamefont {Levy}},
  \bibinfo {author} {\bibfnamefont {Z.}~\bibnamefont {Shen}}, \bibinfo {author}
  {\bibfnamefont {A.~E.}\ \bibnamefont {Schramm}}, \bibinfo {author}
  {\bibfnamefont {Z.}~\bibnamefont {Lu}}, \ and\ \bibinfo {author}
  {\bibfnamefont {P.}~\bibnamefont {Kara}},\ }\href {\doibase
  10.1038/nature17965} {\bibfield  {journal} {\bibinfo  {journal} {Nature}\
  }\textbf {\bibinfo {volume} {534}},\ \bibinfo {pages} {378} (\bibinfo {year}
  {2016})}\BibitemShut {NoStop}%
\bibitem [{\citenamefont {Peppiatt}\ \emph {et~al.}(2006)\citenamefont
  {Peppiatt}, \citenamefont {Howarth}, \citenamefont {Mobbs},\ and\
  \citenamefont {Attwell}}]{peppiatt_bidirectional_2006}%
  \BibitemOpen
  \bibfield  {author} {\bibinfo {author} {\bibfnamefont {C.~M.}\ \bibnamefont
  {Peppiatt}}, \bibinfo {author} {\bibfnamefont {C.}~\bibnamefont {Howarth}},
  \bibinfo {author} {\bibfnamefont {P.}~\bibnamefont {Mobbs}}, \ and\ \bibinfo
  {author} {\bibfnamefont {D.}~\bibnamefont {Attwell}},\ }\href {\doibase
  10.1038/nature05193} {\bibfield  {journal} {\bibinfo  {journal} {Nature}\
  }\textbf {\bibinfo {volume} {443}},\ \bibinfo {pages} {700} (\bibinfo {year}
  {2006})}\BibitemShut {NoStop}%
\bibitem [{\citenamefont {Chang}\ \emph {et~al.}(2017)\citenamefont {Chang},
  \citenamefont {Tu}, \citenamefont {Baek}, \citenamefont {Pietersen},
  \citenamefont {Liu}, \citenamefont {Savage}, \citenamefont {Hwang},
  \citenamefont {Hsiai},\ and\ \citenamefont {Roper}}]{chang_optimal_2017}%
  \BibitemOpen
  \bibfield  {author} {\bibinfo {author} {\bibfnamefont {S.-S.}\ \bibnamefont
  {Chang}}, \bibinfo {author} {\bibfnamefont {S.}~\bibnamefont {Tu}}, \bibinfo
  {author} {\bibfnamefont {K.~I.}\ \bibnamefont {Baek}}, \bibinfo {author}
  {\bibfnamefont {A.}~\bibnamefont {Pietersen}}, \bibinfo {author}
  {\bibfnamefont {Y.-H.}\ \bibnamefont {Liu}}, \bibinfo {author} {\bibfnamefont
  {V.~M.}\ \bibnamefont {Savage}}, \bibinfo {author} {\bibfnamefont {S.-P.~L.}\
  \bibnamefont {Hwang}}, \bibinfo {author} {\bibfnamefont {T.~K.}\ \bibnamefont
  {Hsiai}}, \ and\ \bibinfo {author} {\bibfnamefont {M.}~\bibnamefont
  {Roper}},\ }\href {\doibase 10.1371/journal.pcbi.1005892} {\bibfield
  {journal} {\bibinfo  {journal} {PLOS Comp. Biol.}\ }\textbf {\bibinfo
  {volume} {13}},\ \bibinfo {pages} {e1005892} (\bibinfo {year}
  {2017})}\BibitemShut {NoStop}%
\bibitem [{\citenamefont {Obrist}\ \emph {et~al.}(2010)\citenamefont {Obrist},
  \citenamefont {Weber}, \citenamefont {Buck},\ and\ \citenamefont
  {Jenny}}]{obrist_red_2010}%
  \BibitemOpen
  \bibfield  {author} {\bibinfo {author} {\bibfnamefont {D.}~\bibnamefont
  {Obrist}}, \bibinfo {author} {\bibfnamefont {B.}~\bibnamefont {Weber}},
  \bibinfo {author} {\bibfnamefont {A.}~\bibnamefont {Buck}}, \ and\ \bibinfo
  {author} {\bibfnamefont {P.}~\bibnamefont {Jenny}},\ }\href {\doibase
  10.1098/rsta.2010.0045} {\bibfield  {journal} {\bibinfo  {journal} {Philos.
  Trans. Roy. Soc. A}\ }\textbf {\bibinfo {volume} {368}},\ \bibinfo {pages}
  {2897} (\bibinfo {year} {2010})}\BibitemShut {NoStop}%
\bibitem [{\citenamefont {Stein}(2012)}]{stein_channels_2012}%
  \BibitemOpen
  \bibfield  {author} {\bibinfo {author} {\bibfnamefont {W.~D.}\ \bibnamefont
  {Stein}},\ }\href@noop {} {\emph {\bibinfo {title} {Channels, Carriers, and
  Pumps: An Introduction to Membrane Transport}}}\ (\bibinfo  {publisher}
  {Academic Press},\ \bibinfo {year} {2012})\BibitemShut {NoStop}%
\bibitem [{\citenamefont {Pries}(1996)}]{pries_biophysical_1996}%
  \BibitemOpen
  \bibfield  {author} {\bibinfo {author} {\bibfnamefont {A.}~\bibnamefont
  {Pries}},\ }\href {\doibase 10.1016/0008-6363(96)00065-X} {\bibfield
  {journal} {\bibinfo  {journal} {Cardiovasc. Res.}\ }\textbf {\bibinfo
  {volume} {32}},\ \bibinfo {pages} {654} (\bibinfo {year} {1996})}\BibitemShut
  {NoStop}%
\bibitem [{\citenamefont {Rubenstein}\ \emph {et~al.}(2015)\citenamefont
  {Rubenstein}, \citenamefont {Yin},\ and\ \citenamefont
  {Frame}}]{Rubenstein:2015}%
  \BibitemOpen
  \bibfield  {author} {\bibinfo {author} {\bibfnamefont {D.}~\bibnamefont
  {Rubenstein}}, \bibinfo {author} {\bibfnamefont {W.}~\bibnamefont {Yin}}, \
  and\ \bibinfo {author} {\bibfnamefont {M.~D.}\ \bibnamefont {Frame}},\
  }\href@noop {} {\emph {\bibinfo {title} {{Biofluid Mechanics}}}},\ An
  Introduction to Fluid Mechanics, Macrocirculation, and Microcirculation\
  (\bibinfo  {publisher} {Academic Press},\ \bibinfo {year} {2015})\BibitemShut
  {NoStop}%
\end{thebibliography}

%



\section*{Supplementary material}

## S1.  DERIVATION OF STEADY-STATE SUPPLY ACROSS A CYLINDRICAL VESSEL WALL

### A.  Formulation of the dynamics in terms of the cross-sectional averaged concentration

We begin with the advection diffusion equation in a single straight cylindrical vessel, where we assume that the fluid flow obeys a Poiseuille profile $U(r)$,

$$\frac{\partial C}{\partial t} + U(r)\frac{\partial C}{\partial z} = \kappa \nabla^2 C. \tag{S1}$$

The absorption through a vessel wall sets the boundary condition to

$$\left.\frac{\partial C}{\partial r}\right|_{r=R} + \gamma C(R) = 0, \tag{S2}$$

where we defined the absorption parameter $\gamma$. In analogy to Taylor's derivation of shear dispersion, we define the cross-sectional averaged variables

$$\bar{C} = \frac{1}{A}\int C \mathrm{d}A.$$

With this definition, we can write the concentration $C$ as sum of the cross-sectional averaged concentration $\bar{C}$ and a deviation term $\tilde{C}$,

$$C = \bar{C} + \tilde{C}.$$

Motivated by this separation, we distinguish between derivatives along and perpendicular to the flow direction. The advection diffusion equation and the boundary condition now read

$$\frac{\partial \bar{C}}{\partial t} + \frac{\partial \tilde{C}}{\partial t} + (\bar{U}+\tilde{U})\frac{\partial \bar{C}}{\partial z} + (\bar{U}+\tilde{U})\frac{\partial \tilde{C}}{\partial z} = \kappa \nabla_\parallel^2 \bar{C} + \kappa \nabla_\parallel^2 \tilde{C} + \kappa \nabla_\perp^2 \tilde{C} \quad \text{and}$$

$$\left.\frac{\partial \tilde{C}}{\partial r}\right|_{r=R} + \gamma\left(\tilde{C}(R) + \bar{C}\right) = 0.$$

We take into account that the average and differentiation commute. Together with $\overline{\tilde{U}} = \overline{\tilde{C}} = 0$, we take the cross-sectional average of the advection diffusion equation and find

$$\frac{\partial \bar{C}}{\partial t} + \bar{U}\frac{\partial \bar{C}}{\partial z} + \overline{\tilde{U}\frac{\partial \tilde{C}}{\partial z}} = \kappa \nabla_\parallel^2 \bar{C} + \overline{\kappa \nabla_\perp^2 \tilde{C}}. \tag{S3}$$

We can calculate the last term of above equation using the boundary condition

$$\overline{\kappa \nabla_\perp^2 \tilde{C}} = \kappa \frac{2}{R^2}\int_0^R \nabla_\perp^2 \tilde{C} r \mathrm{d}r = -\frac{2\kappa\gamma}{R}(\bar{C}+\tilde{C}(R)).$$

In Eq. (S3), we only need to estimate the term $\overline{\tilde{U}\partial_z \tilde{C}}$. For this, we subtract the cross-sectional averaged advection diffusion equation from Eq. (S1) and find

$$\frac{\partial \tilde{C}}{\partial t} + \tilde{U}\frac{\partial \bar{C}}{\partial z} + (\bar{U}+\tilde{U})\frac{\partial \tilde{C}}{\partial z} - \overline{\tilde{U}\partial_z \tilde{C}} = \kappa \nabla^2 \tilde{C} + \frac{2\kappa\gamma}{R}(\bar{C}+\tilde{C}(R)).$$

We next employ three approximations, which are in line with reducing the dynamics to center manifold dynamics. The first assumption is that the timescale of diffusion is much smaller than the time scale of advection through the vessel $R^2/\kappa \ll L/U$. This reduces the equation to

$$\tilde{U}\frac{\partial \bar{C}}{\partial z} + (\bar{U}+\tilde{U})\frac{\partial \tilde{C}}{\partial z} - \overline{\tilde{U}\partial_z \tilde{C}} = \kappa \nabla^2 \tilde{C} + \frac{2\kappa\gamma}{R}(\bar{C}+\tilde{C}(R)).$$

Next, we assume that the cross-sectional variation are small $\bar{C} \gg \tilde{C}$. As a high absorption parameter implies a large concentration gradient within the cross-section we demand that the length-scale of advection is much bigger than the cross-section $\gamma R \ll 1$. This implies

$$\tilde{U}\frac{\partial \bar{C}}{\partial z} = \kappa \nabla^2 \tilde{C} + \frac{2\kappa\gamma}{R}(\bar{C} + \tilde{C}(R)).$$

In a last step, we assume that the variation of $\tilde{C}$ is much bigger in radial direction than in flow direction $\partial_r^2 \tilde{C} \gg \partial_z^2 \tilde{C}$. Hence, the length-scale of diffusion along the vessel is much larger than the length-scale of diffusion over the vessel's cross-section, what long slender vessels imply.

$$\tilde{U}\frac{\partial \bar{C}}{\partial z} = \kappa \nabla_r^2 \tilde{C} + \frac{2\kappa\gamma}{R}(\bar{C} + \tilde{C}(R)). \tag{S4}$$

For a known flow profile - here we assumed a Poiseuille flow profile - we can solve the above equation as differential equation for $\tilde{C}$ in dependence of $r$ and $\bar{C}$.

$$\bar{U}\left(1 - \frac{2r^2}{R^2}\right)\frac{\partial \bar{C}}{\partial z} = \kappa \nabla_r^2 \tilde{C} + \frac{2\kappa\gamma}{R}(\bar{C} + \tilde{C}(R)).$$

$$\frac{1}{\kappa}\left(\bar{U}\partial_z\bar{C} - \frac{2\kappa\gamma}{R}(\bar{C} + \tilde{C}(R)) - \bar{U}\frac{2}{R^2}\partial_z\bar{C}r^2\right) = \nabla_r^2 \tilde{C}$$

$$\alpha_1 - \alpha_2 r^2 = \frac{1}{r}\partial_r(r\partial_r\tilde{C}).$$

We find through simple integration

$$\frac{1}{2}\alpha_1 r^2 - \frac{1}{4}\alpha_2 r^4 + \alpha_3 = r\partial_r\tilde{C}.$$

We can immediately determine that $\alpha_3$ yields a logarithmic term in the next integration. Since we demand that the concentration is finite over the whole cross-section, we find that the integration constant $\alpha_3 = 0$. In the next integration step, we find

$$\frac{1}{4}\alpha_1 r^2 - \frac{1}{16}\alpha_2 r^4 + \alpha_4 = \tilde{C}.$$

We fix the value of $\alpha_4$ by accounting for $\overline{\tilde{C}} = 0$.

$$\int_0^R \left(\frac{1}{4}\alpha_1 r^3 - \frac{1}{16}\alpha_2 r^5 + \alpha_4 r\right)\mathrm{d}r = \frac{\alpha_1 R^4}{16} - \frac{\alpha_2 R^6}{16 \cdot 6} + \frac{\alpha_4 R^2}{2},$$

which fixes the value

$$\alpha_4 = \frac{\alpha_2 R^4}{48} - \frac{\alpha_1 R^2}{8}.$$

We find that $\tilde{C}(R)$ is given by

$$\tilde{C}(R) = \frac{\bar{U}R^2}{24\kappa}\left(1 + \frac{\gamma R}{4}\right)^{-1}\partial_z\bar{C} - \frac{\gamma R/4}{1 + \gamma R/4}\bar{C}.$$

Equipped with this result, we can now easily determine the expression $\overline{\tilde{U}\partial_z\tilde{C}}$.

$$\overline{\tilde{U}\partial_z\tilde{C}} = \frac{2\bar{U}}{R^2}\int_0^R \left(\frac{1}{4}\partial_z\alpha_1 r^3 - \frac{1}{16}\partial_z\alpha_2 r^5 + \partial_z\alpha_4 r\right) - \frac{2r^2}{R^2}\left(\frac{1}{4}\partial_z\alpha_1 r^3 - \frac{1}{16}\partial_z\alpha_2 r^5 + \partial_z\alpha_4 r\right)\mathrm{d}r,$$

$$= \frac{2\bar{U}}{R^2}\left[\frac{R^4\partial_z\alpha_1}{16} - \frac{R^6\partial_z\alpha_2}{96} + \frac{R^2\partial_z\alpha_4}{2} - \frac{2R^6\partial_z\alpha_1}{24R^2} + \frac{2R^8\partial_z\alpha_2}{16 \cdot 8 \cdot R^2} - \frac{2R^4\partial_z\alpha_4}{4R^2}\right]$$

$$= -\frac{R^2\bar{U}\partial_z\alpha_1}{24} + \frac{R^4\bar{U}\partial_z\alpha_2}{96}$$

Inserting all definitions and sorting for orders in $\partial_z^n \bar{C}$, we find

$$\frac{\partial \bar{C}}{\partial t} = -\frac{2\kappa}{R^2}\frac{4\gamma R}{4 + \gamma R}\bar{C} - \frac{12 + 5\gamma R}{12 + 3\gamma R}\bar{U}\frac{\partial \bar{C}}{\partial z} + \left(\kappa + \frac{12 + \gamma R}{12 + 3\gamma R}\frac{\bar{U}^2 R^2}{48\kappa}\right)\frac{\partial^2 \bar{C}}{\partial z^2}. \tag{S5}$$

### B. Solution of the cross-sectional averaged dynamics

We now proceed by solving the above partial differential equation for steady state, and we find

$$0 = a\bar{C} + b\partial_z\bar{C} + c\partial_z^2\bar{C}.$$

Making use of an exponential ansatz $\bar{C} = C_0 \exp(-\beta' z)$, we find that $\beta'$ is given by

$$\beta' = \frac{b \pm \sqrt{b^2 - 4ac}}{2c} = \frac{b}{2c}\left(1 - \sqrt{1 - 4\frac{ac}{b^2}}\right),$$

where we used in the last step that the solution has to converge. We define now the parameter $\beta = \beta' \cdot L$, such that the solution is given by $\bar{C} = C_0 \exp(-\beta z/L)$. We then approximate the two coefficients in $\beta$ making use of the above approximation $\gamma R \ll 1$ and find

$$\frac{b}{2c}L \approx \frac{\bar{U}}{2\left(\kappa + \frac{\bar{U}^2 R^2}{48\kappa}\right)}$$

$$= \frac{24\frac{\bar{U}L}{\kappa}}{48 + \frac{\gamma^2 L^2}{\kappa^2 \gamma^2 L^2}} = \frac{24 \cdot \mathrm{Pe}}{48 + \frac{\alpha^2}{S^2}}.$$

We will find that it's useful to sort the terms in the three non-dimensional variables Pe, $S$ and $\alpha$. Here Pe is the Peclet number with $\mathrm{Pe} = \bar{U}L/\kappa$. $S$ is in the style of a Damköhler number $S = \kappa\gamma L/R\bar{U}$. $\alpha$ is giving the ratio of the length-scale of absorption and the length-scale of the vessel $\alpha = \gamma L$. Likewise we find for the second coefficient in $\beta$ again using $\gamma R \ll 1$

$$1 - 4\frac{ac}{b^2} = 1 + 4\frac{\frac{2\kappa}{R^2}\gamma R\left(\kappa + \frac{\bar{U}^2 R^2}{48\kappa}\right)}{\bar{U}^2} = 1 + 8\frac{\gamma R\kappa^2}{R^2\bar{U}^2} + \frac{\gamma R}{6} = 1 + \frac{8S}{\mathrm{Pe}} + \frac{\alpha^2}{6\mathrm{Pe}S}.$$

In the context of blood flow, it becomes useful not to express $\bar{C}$ in terms of an initial concentration at a vessel entrance $C_0$, but with an initial flux into the vessel. The flux is defined with contribution from the advective and the diffusive flux. We thus write

$$J = \bar{U}\bar{C} - \kappa\partial_z\bar{C}.$$

At the beginning of the vessel, where $z = 0$, we find the relation

$$C_0 = \frac{J_0}{\bar{U} + \kappa\frac{\beta}{L}}.$$

Finally, we define the supply $\phi$ as concentration flux through the surface of the vessel wall and thus

$$\phi = -\int_S \kappa \nabla C \mathrm{d}a.$$

Making use of the boundary condition in Eq. (S2), we find

$$\phi = -\int_S \kappa \nabla C \mathrm{d}a = -\kappa \int_S \left.\frac{\partial C}{\partial r}\right|_{r=R} \mathrm{d}a = 2\pi\kappa\gamma R \int_0^L (\bar{C} + \tilde{C}(R))\mathrm{d}z. \tag{S6}$$

Inserting all definitions, we find for the supply

$$\phi = J_0 \frac{1}{1 + \frac{\beta}{\mathrm{Pe}}} \cdot \left(\frac{\frac{\alpha^2}{12S\mathrm{Pe}} + 2\frac{S}{\beta}}{1 + \frac{\alpha^2}{4S\mathrm{Pe}}}\right) \cdot \left(1 - \exp\left(-24 \cdot \mathrm{Pe} \cdot \frac{\sqrt{1 + \frac{8S}{\mathrm{Pe}} + \frac{\alpha^2}{6\mathrm{Pe}S}} - 1}{48 + \frac{\alpha^2}{S^2}}\right)\right). \tag{S7}$$

This lengthy expression is unintuitive at first sight. In the manuscript we mapped out the dynamics of Eq. (S7). Though this expression is lengthy, one can easily apply the approximations we use in the manuscript to define the three different regimes in terms of Pe and $S$.

### C. Approximation of the supply expression

We now proceed to estimate how big different terms are for the different regimes. We start by focusing on the Taylor approximations conditions that need to be fulfilled for all regimes

$$R \ll L \implies \frac{\alpha}{S\text{Pe}} \ll 1, \tag{S8}$$

$$\gamma R \ll 1 \implies \frac{\alpha^2}{S\text{Pe}} \ll 1, \tag{S9}$$

$$\frac{R^2}{\kappa} \ll \frac{L}{U} \implies \frac{\alpha^2}{S^2\text{Pe}} \ll 1 \implies \frac{\alpha^2}{S^2} \ll \text{Pe}. \tag{S10}$$

#### 1. Advective regime

We next approximate the supply in the different transport regimes. We start with the advective regime, where $S \ll 1$ and $S \ll \text{Pe} \to S/\text{Pe} \ll 1$. We first approximate the bracket expression in $\beta$ where we find using $S/\text{Pe} \ll 1$

$$\left(\sqrt{1 + \frac{8S}{\text{Pe}} + \frac{\alpha^2}{6\text{Pe}S}} - 1\right) \longrightarrow \frac{1}{2}\frac{\alpha^2}{6\text{Pe}S}, \tag{S11}$$

as we series expanded the square root expression for small values of $\frac{\alpha^2}{S\text{Pe}} \ll 1$. For the prefactor of the bracket, we find since $S \ll 1$

$$\frac{24 \cdot \text{Pe}}{48 + \frac{\alpha^2}{S^2}} \longrightarrow \frac{24\text{Pe}}{\frac{\alpha^2}{S}}. \tag{S12}$$

As a result, $\beta$ is given by

$$\beta(\text{Pe}, S, \alpha) \longrightarrow 2S. \tag{S13}$$

Note that $2S \ll 1$, which implies that $\beta$ is small for the advective regime.

#### 2. Diffusive regime

We next focus on the diffusive regime, where $\text{Pe}S \ll 1$ and $\text{Pe} \ll S \to S/\text{Pe} \gg 1$. This implies $\text{Pe} \ll 1$. Note, that this further implies with the Taylor approximations

$$\frac{\alpha^2}{S^2} \ll \text{Pe} \ll 1. \tag{S14}$$

Focusing now on the prefactor of the bracket expression, we find

$$\frac{24 \cdot \text{Pe}}{48 + \frac{\alpha^2}{S^2}} \longrightarrow \frac{\text{Pe}}{2}. \tag{S15}$$

For the bracket expression, we find using $S/\text{Pe} \gg 1$

$$\left(\sqrt{1 + \frac{8S}{\text{Pe}} + \frac{\alpha^2}{6\text{Pe}S}} - 1\right) \longrightarrow \sqrt{\frac{8S}{\text{Pe}}}. \tag{S16}$$

As a result, $\beta$ is given by

$$\beta(\text{Pe}, S, \alpha) \longrightarrow \sqrt{2\text{Pe}S}. \tag{S17}$$

Note that $\text{Pe}S \ll 1$, which implies that $\beta$ is small for the diffusive regime.

### 3. All absorbing regime

We next focus on the all absorbing regime, where $S \ll 1$ and $\mathrm{Pe}S \gg 1$. With this we find that

$$\frac{24 \cdot \mathrm{Pe}}{48 + \frac{\alpha^2}{S^2}} \left( \sqrt{1 + \frac{8S}{\mathrm{Pe}} + \frac{\alpha^2}{6\mathrm{Pe}S}} - 1 \right) \longrightarrow \frac{\mathrm{Pe}}{2} \left( \sqrt{1 + \frac{8S}{\mathrm{Pe}}} - 1 \right). \tag{S18}$$

We next distinguish the cases $S/\mathrm{Pe} \gg 1$ and $S/\mathrm{Pe} \ll 1$. Starting with $S/\mathrm{Pe} \gg 1$, we find

$$\beta(\mathrm{Pe}, S, \alpha) \longrightarrow \sqrt{2\mathrm{Pe}S}, \tag{S19}$$

where $\beta$ is big due to $\mathrm{Pe}S \gg 1$. For $S/\mathrm{Pe} \ll 1$, we find

$$\beta(\mathrm{Pe}, S, \alpha) \longrightarrow 2S \tag{S20}$$

where $\beta$ is big due to $S \gg 1$.

$$\beta(\mathrm{Pe}, S, \alpha) = \frac{24 \cdot \mathrm{Pe}}{48 + \frac{\alpha^2}{S^2}} \left( \sqrt{1 + \frac{8S}{\mathrm{Pe}} + \frac{\alpha^2}{6\mathrm{Pe}S}} - 1 \right) \tag{S21}$$

## S2. DERIVATION OF THE ABSORPTION PARAMETER $\gamma$

To our knowledge no measurements of the absorption parameter $\gamma$ have yet been conducted. To estimate values for the absorption parameters either a *microscopic* approach or a *macroscopic* approach can be used. For the microscopic approach we estimate an absorption parameter using permeabilities through lipid bilayers, channels conductance, channel abundance and other physiological features allowing the extraction of solute from the blood vessel through the vessel wall. Though there are experiments assessing some of these parameters, see e.g. [1–3], to conclude on the absorption parameter using the microscopic approach demands a large number of additional assumptions which are not validated by in vivo experiments. Instead we use here a macroscopic approach. For the macroscopic approach we estimate the absorption parameter using the model developed in Supplemental Material S1 and the macroscopic observable of the overall measured absorption in the brain. This approach allows us to derive a more physiological estimate of the absorption parameters, as we rely on *in vivo* experiments and make less assumptions compared to the microscopic approach.

Leybaert et al. [4] measured the total absorption of glucose in the rat brain to be 10%. We assume that the main absorption happens in capillary beds and thus between penetrating ateriol and penetrating venule and that the number $n$ of vessels between ateriol and venule is of the order $n = \mathcal{O}(10)$. In the data set considered by us [5] the median vessel radius is $R = 2\,\mu\mathrm{m}$, vessel length $L = 50\,\mu\mathrm{m}$, and flow velocity $U = 400\,\mu\mathrm{m\,s^{-1}}$. For the estimate here, we consider the total supply and thus assume an effective length of $L_\mathrm{tot} = nL$. We estimate the absorption parameter making use of the relation

$$\frac{C_\mathrm{abs}}{C_\mathrm{tot}} = 1 - e^{-\beta}. \tag{S22}$$

We solve this equation for $\gamma$, where we take into account that $\gamma R \ll 1$. We find with the above equation and the median parameters of the vascular network a value for the absorption parameter of the order $\gamma = \mathcal{O}(100\,\mathrm{m}^{-1})$. Here, we chose $\gamma = 200\,\mathrm{m}^{-1}$. We chose a value on the upper boundary of the estimate to increase the statistics in the diffusive and all absorbing regime. Note that with this choice of $\gamma$, we still fulfill the assumption $\gamma R \ll 1$. Furthermore note, that the robust results are independent of the exact value of $\gamma$, compare with the figures in Supplemental Material S6.

## S3. ESTIMATION OF UPSTREAM CHANGES ON $C_0$

We find that for the advective and the diffusive regime, we can estimate the supply by

$$\phi_\mathrm{advective} \approx \phi_\mathrm{diffusive} \approx 2\pi R L \cdot \kappa \gamma \cdot C_0,$$

which renders supply independent of the flow velocity. For upstream vessels in the advective regime, the solute flux exiting a vessel is given by

$$J_{\text{up,out}} = J_{0,\text{up}} - \phi_{\text{up}}. \tag{S23}$$

This is the solute flux that enters in Eq. (14) as influx into a network node. We rewrite the equation above in terms of the initial concentration $C_{0,\text{up}}$, which is the initial concentration at the beginning of the upstream vessel, to find

$$J_{\text{up,out}} = C_{0,\text{up}} \cdot \pi R^2 U - C_{0,\text{up}} \cdot 2\pi R L \cdot \kappa \gamma, \tag{S24}$$

where flow velocity, radius, and length of the upstream vessel were used. With this and the supply through the upstream vessel we can now compute the solute outflux of the upstream vessel as

$$J_{\text{up,out}} = C_{0,\text{up}} \cdot \pi R^2 U \left(1 - \frac{2\kappa\gamma L}{RU}\right) = C_{0,\text{up}} \cdot \pi R^2 U \left(1 - 2S\right). \tag{S25}$$

Thus, we find that $C_{0,\text{up}}$ and hence $C_{\text{in}}$ in Eq. (14) change by the factor $(1 - 2S_{\text{new}})/(1 - 2S)$. This factor is function of $\Delta U$. Taking the Taylor series to first order in $\Delta U$, we find that $C_{\text{in,dil}}$ is changed by a factor

$$C_{\text{in,dil}} = C_{\text{in}} \left(1 + \frac{2S}{1 - 2S}\Delta U + \mathcal{O}\left((\Delta U)^2\right)\right). \tag{S26}$$

We previously identified the advective regime to be characterized by $S \ll 1$. And even as we cannot estimate the precise expression of $\Delta U$ without knowing the exact network topology, it is reasonable to assume that the change for a vessel dilation of 10% is below 100% for the flow velocity, what implies $\Delta U < 1$. As a consequence, we find that the contribution of of upstream effects on $C_{\text{in}}$ and hence $C_0$ in Eq. (15) is negligible, which is also in line with the simulation results, see for this the figure in the Supplemental Material S5.

For one of the outflow tubes being in the diffusive regime, while the inlet tube and at least one other inflow tube in the advective regime, we need to focus on the denominator of Eq. (14). Note, that in the data set considered here, vessel branching of have a similar radius. We find that a vessel in the diffusive regime have a low flow $Q$. The size of the diffusive flux is approximately the same for vessels in the diffusive regime and vessels in the advective regime. For the denominator of Eq. (14) this implies that the advective flux terms dominate the denominator as long as at least on of the outflow vessels is in the advective regime. With the same reasoning as before, we can thus neglect all diffusive terms in the equation. If the upstream inlet vessel is in the diffusive regime, all vessels branching off must also be in the diffusive regime. For this case, the approximation above fails and the denominator increases as one of the vessels is dilated.

Focusing on the nominator of Eq. (14) and how $J_{\text{up,out}}$ changes for an upstream inlet vessel in the diffusive regime, we find the analogous expression

$$J_{\text{up,out}} = C_{0,\text{up}} \cdot \pi R^{\frac{3}{2}} \sqrt{2\gamma} \kappa \left(1 - \sqrt{2S\text{Pe}}\right). \tag{S27}$$

Thus, for a vessel in the diffusive regime $C_{\text{in}}$ is truly independent of upstream effects as this expression is fully independent of $U$. Thus a vessel dilation will cause no upstream effects. The nominator of Eq. (14) will not change due to a vessel dilation, leaving $C_{\text{in,dil}}$ with a small decrease.

## S4. SPATIAL CORRELATION OF SUPPLY IN DOWNSTREAM VESSELS

We find that the typical response of a dilating vessel in both the advective and the diffusive regime is to increase supply downstream at the cost of reducing supply in the parallel branch. Can we understand how spatial correlations in downstream vessels arise without knowing the exact network topology?

For this we focus again first at a dilating vessel in the advective regime. Following the same line of reasoning as in the Supplemental Material S3, we understand the dilated tube as inlet tube for vessels that are directly downstream the dilated vessel. In contrast to the calculations in the Supplemental Material S3, here the tube itself is dilated by a factor $\Delta R$. To first order in $\Delta U$, we find that $C_{\text{in,dil}}$ is altered by the factor.

$$C_{\text{in,dil}} = C_{\text{in}} \left(\frac{(1+\Delta R)^3 - 2S}{(1+\Delta R)(1-2S)} + \frac{2S}{(1-2S)(1+\Delta R)}\Delta U + \mathcal{O}\left((\Delta U)^2\right)\right). \tag{S28}$$

With

$$\frac{(1+\Delta R)^2 - 2S}{(1+\Delta R)(1-2S)} > 1 \quad \text{for } S \ll 1,$$

which is true for vessels in the advective regime. The second term can be positive or negative depending on the sign of $\Delta U$. To predict the value of $\Delta U$, the full network topology must be known. Furthermore, non-linear corrections for the hematocrit affect $\Delta U$. Nevertheless, we can estimate $\Delta U$ with a rule of thumb. For this, we assume a constant pressure drop over the dilating vessel and neglect correction for the hematocrit. In this case, the change in the velocity scales like $\Delta U \propto (\Delta R)^2$ and the second term is also positive. More solute is drawn into the dilated branch than is consumed more by increased absorption along the dilated vessel.

Further downstream effects follow the same form as described in the Supplemental Material S5, where we find

$$C_{\text{in,dil}} = C_{\text{in}} \left(1 + \frac{2S}{1-2S}\Delta U + \mathcal{O}\left((\Delta U)^2\right)\right).$$

As for the dilated tube, whether the correction term increases or decreases the further supply downstream depends on the sign of $\Delta U$.

In line with the argument in Supplemental Material S5, the further correction terms downstream the dilated vessel are small, as $\Delta U$ is small. Note, however that there is a topological difference between vessels upstream and vessels downstream a dilated vessel. The consequence of the topological difference can be understood drawing the analogy to circuit networks. Recall, that the microvasculature has a loopy topology. For the further argument, we focus on the loop in which the dilated vessel is located. Recall now, that the flow through the branch of a loop is determined by the total resistance along the branch. The change in resistance is strongest in the branch in which the dilated vessel is located. However, for the upstream vessels, the total path through the vasculature must be considered. Here, the total change in resistance is buffered by the parallel branches leaving a smaller change in the flow. As a result $\Delta U$ is bigger in the downstream vessels than in the upstream vessel.

A reduction of supply in the vessels of the parallel branches can be understood along the same lines. Decreasing the resistance along the branch with the dilated vessel, more flow - and hence more advective flux - is drawn into the branch with the dilated tube. As this is on expense of the the flux in the parallel branches, the supply in the parallel branches is decreased.

The argument is easier for vessels in the diffusive regime. Here, the supply in downstream vessels is increased as the influx scales with $(\Delta R)^{3/2}$, while the increased supply along the dilated vessel scales with $\Delta R$. Thus the net outflux exiting the dilated tube is also increased.

Note that on the one hand, the effect of altered supply downstream is weak on the level of individual vessels, but correlations sum up the effect in a spatial region of the microvsasculature and by this increase the effect. Note on the other hand, that changes in downstream and parallel vessels strongly depend on the network topology and that $\Delta U$ is subject to non-linear corrections due to the hematocrit. An increase in supply downstream the dilated vessel and a decrease in supply in vessels parallel to the dilated vessel can thus only be considered as a rule of thumb, in line with Fig. 3 (b). To estimate the full effect the exact network topology of the microvasculature must be known.

[1] Joao M. N. Duarte, Florence D. Morgenthaler, Hongxia Lei, Carol Poitry-Yamate, and Rolf Gruetter, "Steady-State Brain Glucose Transport Kinetics Re-Evaluated with a Four-State Conformational Model," Front. Neuroenerg. **1**, 1–10 (2009).
[2] H. Fischer, R. Gottschlich, and A. Seelig, "Blood-Brain Barrier Permeation: Molecular Parameters Governing Passive Diffusion," J. Membrane Biol. **165**, 201–211 (1998).
[3] Warren L. Lee and Amira Klip, "Shuttling glucose across brain microvessels, with a little help from GLUT1 and AMP kinase. Focus on AMP kinase regulation of sugar transport in brain capillary endothelial cells during acute metabolic stress," Am. J. Physiol. Cell Physiol. **303**, C803–C805 (2012).
[4] Luc Leybaert, "Neurobarrier Coupling in the Brain: A Partner of Neurovascular and Neurometabolic Coupling?" J. Cereb. Blood Flow Metab. **25**, 2–16 (2005).
[5] Pablo Blinder, Philbert S. Tsai, John P. Kaufhold, Per M. Knutsen, Harry Suhl, and David Kleinfeld, "The cortical angiome: an interconnected vascular network with noncolumnar patterns of blood flow," Nat. Neurosci. **16**, 889–897 (2013).

## S5. CHANGE IN $C_0$ IN SIMULATIONS

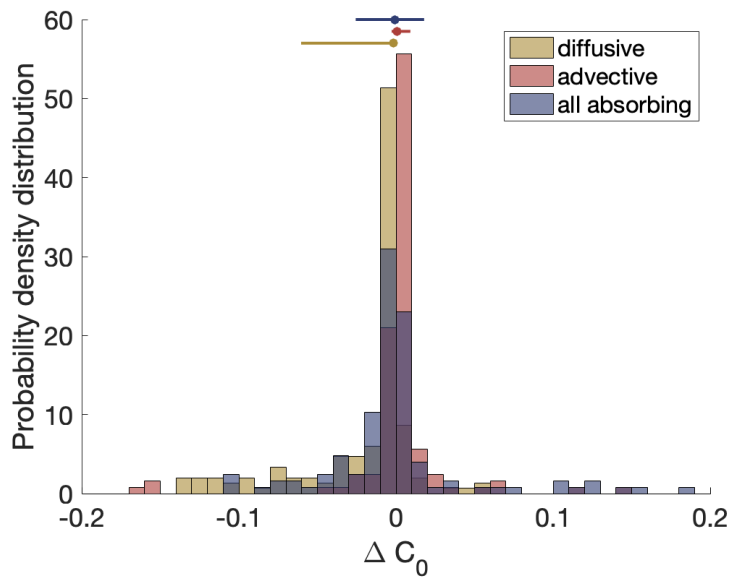

FIG. S1. Histogram of change in initial concentration $C_0$ due to a single vessel dilating by 10%. The advective regime shows no change in the initial concentration $C_0$ at the beginning of a vessel that is dilated. Also, vessels in the diffusive regime show little response for single vessel dilation, with more vessels showing a decrease in $C_0$ compared to vessels in the advective regime. For this histogram the same simulations as shown in Fig. 2 are evaluated. Lines indicate a range covering 69% with both a percentage of 15.5% showing a lower or higher change in $C_0$ outside the indicated range. For each histogram 120 random vessel of the respective regime were randomly chosen and dilated.

## S6.   CHANGE IN SUPPLY FOR ALTERNATIVE CHOICES OF $\gamma$ AND $\kappa$ AND DIFFERENT NETWORK BOUNDARY CONDITIONS

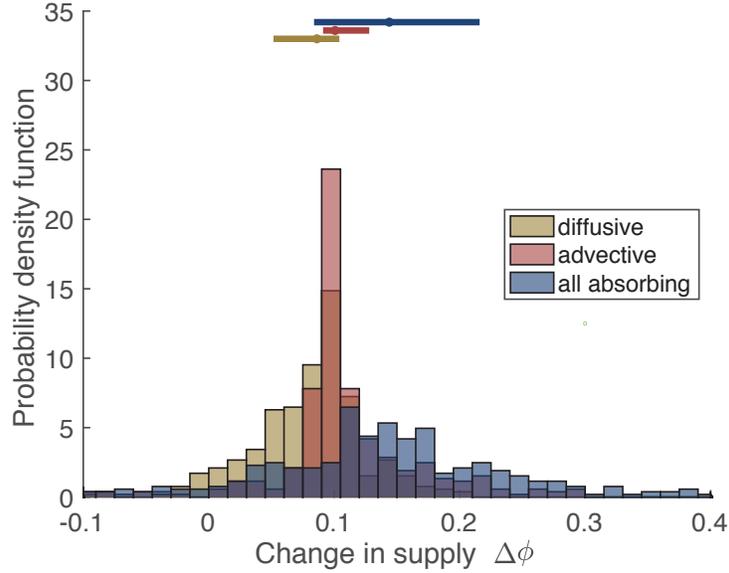

FIG. S2. The advective regime is robust in increasing supply by dilation also for an alternative parameter set than shown in Fig. 2. Here, an diffusivity of $\kappa = 1 \times 10^{-8}\,\mathrm{m^2\,s^{-1}}$ and an absorption parameter of $\gamma = 200\,\mathrm{m^{-1}}$ are chosen. Histogram of change in supply $\Delta\phi$ due to a single vessel dilating by 10%. Lines span from the first quartile of the data to the third quartile. Big dots indicate the median. For each histogram 450 random vessel of the respective regime were randomly chosen and dilated.

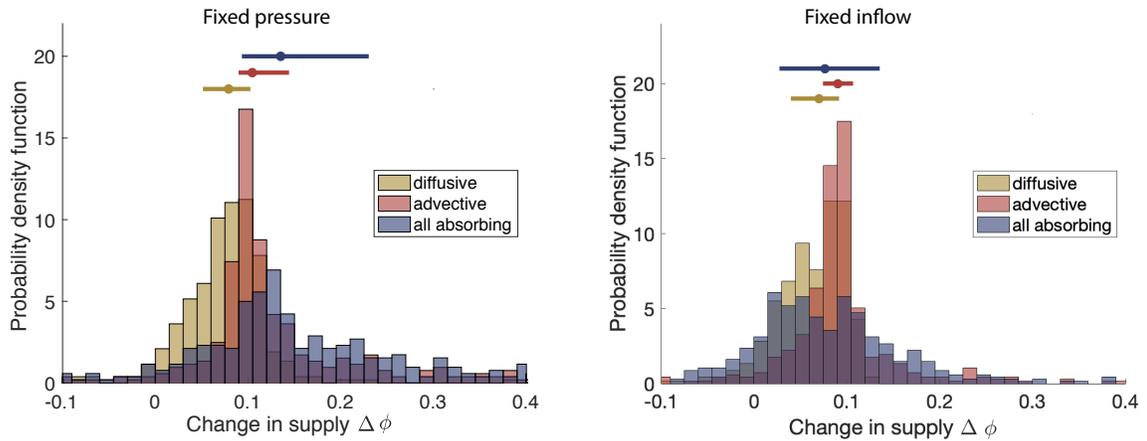

FIG. S3. The advective regime is robust in increasing supply by dilation also for an alternative parameter set than shown in Fig. 2 and for a fixed inflow boundary condition. Here, an diffusivity of $\kappa = 1 \times 10^{-8}\,\mathrm{m^2\,s^{-1}}$ and an absorption parameter of $\gamma = 800\,\mathrm{m^{-1}}$ are chosen. This choice of parameters corresponds to centering the data around Pe = 1 and $S = 1$ yielding the most equal distribution of data in the three identified transport regimes. On the left, nodes at the boundary of the network were held at fixed pressure as single vessels were dilated. On the right, tubes starting from the boundary were held at a fixed inflow $Q$ as single vessels werde dilated. Histogram of change in supply $\Delta\phi$ due to a single vessel dilating by 10%. Lines span from the first quartile of the data to the third quartile. Big dots indicate the median. For each histogram 450 random vessel of the respective regime were randomly chosen and dilated. A smaller network excerpt compared to Fig. 2 was chosen, which contains only 3340 vessels.

## S7. IMPACT OF HEMATOCRIT ON ROBUSTNESS

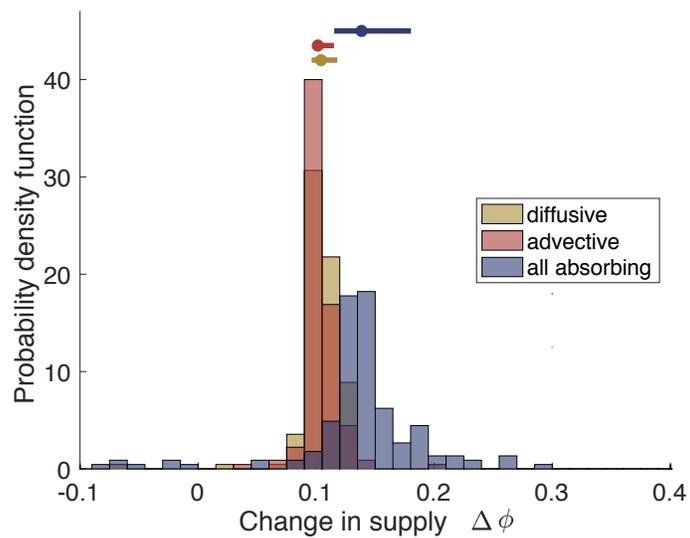

FIG. S4. Histogram of change in supply $\Delta\phi$ due to a single vessel dilating by 10% for the same parameters as Fig. 2 but neglecting the effect of hematocrit when calculating the flows within the networks shows no difference to original Fig. 2. Lines span from the first quartile of the data to the third quartile. Big dots indicate the median. For each histogram 450 random vessel of the respective regime were randomly chosen and dilated

## S8. DEVIATION FROM ROBUSTNESS

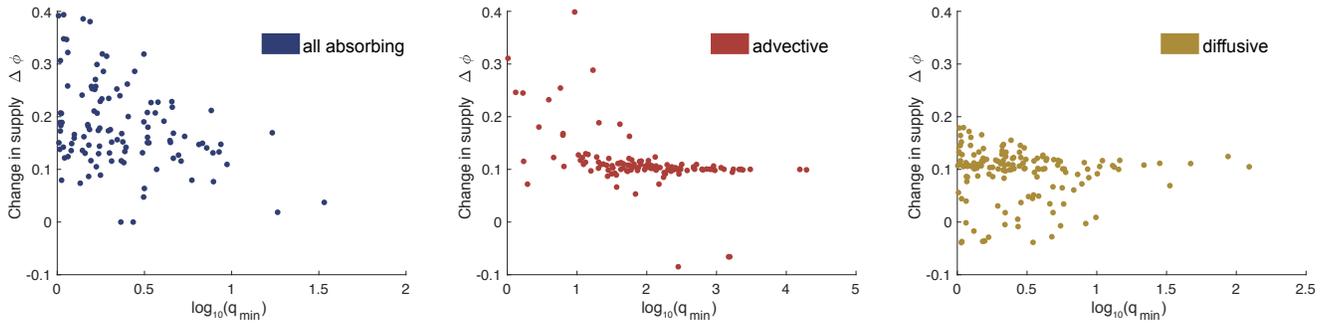

FIG. S5. The quality of the theoretical prediction is decreased close to the border of the regimes. Mapped out on the y-axis is the change in supply displayed in the histogram in Fig. 2. Mapped out on the x-axis is the distance to the border of the regime identified in Fig. 1. $q_{\min}$ is a fold change needed to let a vessel change the regime. Intuitively, $\log_{10}(q_{\min})$ can be understood as the minimal distance of each data point to a regime border in the logarithmic space, i.e. distance of point from the border in Fig. 1. Notably vessels far in the advective regime display a 10% increase in supply while vessels close to the intersection with the other regimes are not as robust in agreement with analytical calculations.

## S9. FLOW CHART FOR CALCULATION OF ABSORPTION PROFILE

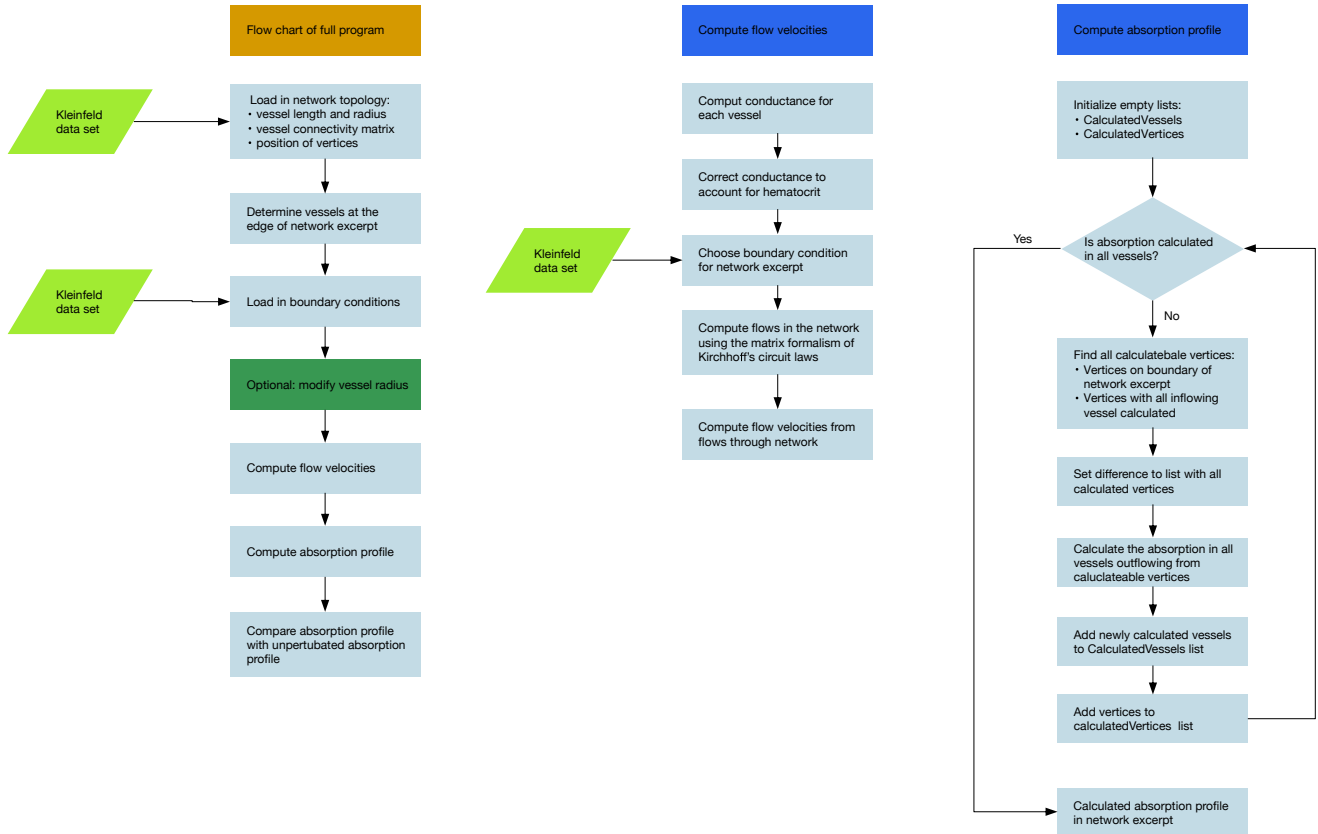

FIG. S6. Flow chart giving an overview over the numerical procedure to compute absorption profiles. On the left, the full outline of the numerical procedure is shown. In the middle, the calculations of flow velocities is visualised, showing how the effect for the hematocrit is accounted for. On the right, the procedure of the calculation of the absorption profiles is shown. We are happy to provide elements or the full code upon request.




\end{document}